\documentclass[apj]{emulateapj}
\usepackage{float}
\usepackage{epsfig}
\usepackage{graphicx}
\usepackage{graphics}
\usepackage[latin1]{inputenc}
\usepackage{latexsym}
\usepackage{footnote}
\usepackage[bottom]{footmisc}
\newcommand{\cl}{\textit{clear}}
\newcommand{\nd}{{\nodata}}

\begin{document}
\submitted{Accepted for Publication in the Astronomical Journal}
\title{Stellar Variability and Flare Rates from Dome A, Antarctica\\ using 2009 and 2010 CSTAR Observations}

\author{Ryan J.~Oelkers\altaffilmark{1,*}, Lucas M.~Macri\altaffilmark{1}, Lifan Wang\altaffilmark{1,2,3}, Michael C.~B.~Ashley\altaffilmark{4}, Xiangqun Cui\altaffilmark{3,5},\\ Long-Long Feng\altaffilmark{2,3}, Xuefei Gong\altaffilmark{3,5}, Jon S.~Lawrence\altaffilmark{4,6}, Liu Qiang\altaffilmark{3,7}, Daniel Luong-Van\altaffilmark{4}, \\Carl R.~Pennypacker\altaffilmark{8}, Xiangyan Yuan\altaffilmark{3,5}, Donald G.~York\altaffilmark{9}, Xu Zhou\altaffilmark{3,7}, Zhenxi Zhu\altaffilmark{2,3}}

\altaffiltext{1}{George P.~and Cynthia W.~Mitchell Institute for Fundamental Physics and Astronomy, Department of Physics and Astronomy Texas A\&M University, College Station, TX 77843, USA}
\altaffiltext{2}{Purple Mountain Observatory, Chinese Academy of Sciences, Nanjing, China}
\altaffiltext{3}{Chinese Center for Antarctic Astronomy, Nanjing, China}
\altaffiltext{4}{School of Physics, University of New South Wales, NSW, Australia}
\altaffiltext{5}{Nanjing Institute of Astronomical Optics and Technology, Nanjing, China}
\altaffiltext{6}{Australian Astronomical Observatory, NSW, Australia}
\altaffiltext{7}{National Astronomical Observatories, Chinese Academy of Sciences, Beijing, China}
\altaffiltext{8}{Institute of Nuclear and Particle Astrophysics, Lawrence Berkeley National Laboratory, Berkley, CA, USA}
\altaffiltext{9}{Department of Astronomy and Astrophysics and Enrico Fermi Institute, University of Chicago, Chicago, IL, USA}
\altaffiltext{*}{Corresponding author, {\tt ryan.oelkers@physics.tamu.edu}}

\begin{abstract}
The Chinese Small Telescope ARray (CSTAR) carried out high-cadence time-series observations of $\sim 20.1$~square degrees centered on the South Celestial Pole during the 2008, 2009 \& 2010 winter seasons from Dome A in Antarctica. The nearly-continuous 6 months of dark conditions during each observing season allowed for $>10^{6}$ images to be collected through \textit{gri} and \cl\ filters, resulting in the detection of $>10^4$ sources over the course of 3 years of operation. The nearly space-like conditions in the Antarctic plateau are an ideal testbed for the suitability of very small-aperture ($<20$~cm) telescopes to detect transient events, variable stars and stellar flares. We present the results of a robust search for such objects using difference image analysis of the data obtained during the 2009 \& 2010 winter seasons. While no transients were found, we detected 29 flaring events and find a normalized flaring rate of $5\pm4\times10^{-7}$ flare/hour for late-K dwarfs, $1\pm1\times10^{-6}$ flare/hour for M dwarfs and $7\pm1\times10^{-7}$ flare/hour for all other stars in our sample. We suggest future small-aperture telescopes planned for deployment at Dome A would benefit from a tracking mechanism, to help alleviate effects from ghosting, and a finer pixel scale, to increase the telescope's sensitivity to faint objects. We find that the light curves of non-transient sources have excellent photometric qualities once corrected for systematics, and are limited only by photon noise and atmospheric scintillation. 

\end{abstract}

\section{Introduction}
The hunt for transiting exoplanets, micro-lensing events and variable stars has accelerated the use of small-aperture telescopes to provide high-cadence time-series photometry during the past three decades. Small-aperture telescopes are well suited to this task because of their lower relative cost, large fields of view (FoV) and reproducibility \citep{Bakos2002, Pollacco2006, Pepper2007, Law2013}. A byproduct of these searches has been a flood of observations and categorization of transient events using small to moderate aperture telescopes \citep{Law2009, Shappee2014, Kessler2015}.

Transient events are typically described as sudden increases in magnitude lasting between a few hours and a few days. The detection of such events is highly nontrivial given the lack of {\it a priori} information on their location and timing. Most of the above-referenced surveys use difference image analysis (DIA) data reduction pipelines to increase transient detection. DIA can be summarized as subtracting a convolved high-quality reference image from fresh science images. Correlated residuals left on the resulting frame (such as a sudden increase in flux at a previously blank location) denote a statistically significant change that can lead to the immediate and robust detection of a transient. The use of DIA has greatly minimized the impact of systematics associated with these searches. Unfortunately, most surveys are constrained in temporal coverage by nature (i.e. the day/night cycle and weather) and other uses of the telescope.

The Chinese Small Telescope ARray~(CSTAR) was designed to test the feasibility and quality of an observatory stationed at Dome A on the Antarctic Plateau ($87.3667^{\circ}$S,~$77.3500^{\circ}$E). Dome A is considered to be one of the most promising observing sites on Earth with low temperature (-60~to~-80$^{\circ}$~C), high altitude (4200~m), extremely stable atmospheric conditions ($<0.4$~mag extinction for 70\% of the time~) and nearly-uninterrupted dark conditions for 6 months \citep{Zou2010, Zhou2010, Wang2011, WangS2012, Wang2013, Zhou2013, WangS2014b,  Oelkers2015}. The combination of these factors allow for astronomical observations to take place in nearly space-like conditions. While the primary objective of the telescope was site testing, the initial three years of operations yielded over $10^6$ unique photometric data points for $>10^4$ astronomical sources in \textit{gri} and \cl\ filters. Several analyses of this data set have yielded a variety of interesting and rare variable stars such as Blazhko-effect RR Lyraes, exoplanet candidates, eccentric eclipsing binaries, pulsating variables in eclipsing systems and a wide variety of irregular variable sources \citep{Zou2010, Zhou2010, Wang2011, Wang2013, WangS2014, Qian2014, Oelkers2015, WangS2015, Yang2015}. 

While the large FoV ($\sim20~$sq. deg) and fast cadence ($20-40$~s) of CSTAR are useful for transient searches, extragalactic events such as type la supernovae should remain elusive given the instrument's shallow limiting magnitudes. SNe la have peak magnitudes $\sim -19.17, -19.03, \& -18.5$~mag in \textit{g,r}\&\textit{i} \citep{Phillips1993, Folatelli2010}. Therefore, a SNe Ia would need to be at a distance $<39.8$~Mpc to be detected in \textit{i} with CSTAR ($<31.1$~Mpc for \textit{g} $\&$ \textit{r}). Using the NASA/IPAC Extragalactic Database, we identified 4 galaxies within the CSTAR FoV that lie at distances closer than 40~Mpc, plus an additional 3 systems with $K<12$ in the 2MASS catalogue \citep{Skrutskie2006}. Given a supernovae rate of $10^{-2}$~yr$^{-1}$ and the 250~d baseline of the 2009-10 CSTAR observations we calculate the effective supernovae rate to be $<0.02$ for a given CSTAR observing season. These probabilities are consistent with the null result described below.

The rapid cadence of CSTAR allows stars to be studied in an unprecedented fashion. Stellar flares and random changes in variability can be investigated under these observing conditions. We expect these events will be the largest contributor to the total number of detected transients given the large number of sources in the CSTAR field with nearly complete temporal coverage. We have chosen to specifically study the flare rate of Late-K and M~dwarfs in addition the field's general flare rate because of the large amount of literature available to compare to. Previous studies of the flaring rate of M-dwarfs have been primarily undertaken using Stripe 82 in the Sloan Digital Sky Survey (SDSS) archival data and the primary Kepler Space Telescope mission data set \citep{York2000, Borucki2010}; these searches were targeted because of the high frequency of observations and the large area of coverage. Studies of Stripe 82 searched over 50,000 M-dwarf light curves for flares and found 271 events leading to a flaring fraction of $0.01-.28\%$ over a $\sim10$ year baseline which is greatly dependent on the distance from the Galactic Plane. These searches found the most active M-dwarfs will flare with a $\Delta$\textit{g}$=1.1$ at ~$1.3$ flares hr$^{-1}$~sq. deg$^{-1}$ \citep{Kowalski2009}. CSTAR provides a unique opportunity to test these rates because it is not constrained in temporal coverage as described above.

An active M-dwarf would need to be $<150$~pc from the Sun to be detected by CSTAR given the limiting magnitude of the system. The CSTAR field is located at \textit{l}$=303^{\circ}$ and \textit{b}$=-27^{\circ}$ and directed towards the Milky Way halo. A simple query of the TriLegal galactic model \citep{Girardi2012} shows there should be $\sim$1600 K/M-dwarfs in the CSTAR field within the magnitude range where CSTAR could detect a flare. Assuming the flaring fraction from above we would expect $1-4$ of these K/M dwarfs would provide a flare. However, given the above flaring rates we would expect $\sim26$ flares/hr in the 20.1 sq. degree FoV of CSTAR. However, due to the short lifetime of such events ($\sim5-45$~min), care will be necessary to distinguish the flare from a brightening event introduced by the detector's systematics.

As described in detail below, CSTAR photometry can have systematics such as daily aliasing, ghost reflections, image defocusing and attenuation due to intermittent lens frosting. This paper presents a comprehensive characterization of all of these effects as part of a search for classical variables, transients and stellar flares in the 2009 and 2010 observations. Our paper is organized as follows: \S2 summarizes the 2009 and 2010 CSTAR observations; \S3 details the data processing steps; \S4 describes our photometric reduction process; \S5 discusses the sources of systematic effects; \S6 details the search for variability, flares and transients; \S7 presents our results.

\section{Observations}

CSTAR was deployed to Dome A in early 2008 and carried out observations during three Antarctic winter seasons supported by the PLATO observatory \citep{Lawrence2009} before being returned to China for comprehensive upgrades in early 2011; the following description applies to the original version of the system. It is composed of four Schmidt-Cassegrain wide-field telescopes, each with a 145mm aperture and a field of view 20.1 sq. degrees. The focal planes contain ANDOR DV435 1K$\times$1K frame-transfer CCDs with a pixel size of 13$\mu$m, equivalent to a plate scale of $15\arcsec/$pix. Filters are mounted at the top of the optical tubes, with a 10W electric current run through a coating of indium tin oxide to prevent frosting \citep{Yuan2008}. Three of the filters are standard SDSS \textit{gri} while the remaining one is a clear filter (hereafter, {\cl}). CSTAR contained no moving parts and the telescopes did not track, so as to minimize possible failures during the observing seasons. The telescopes are pointed towards the SCP and exposures are short (5-40s) to keep the resulting drift from subtending more than a pixel. The telescopes operated unattended with power supplied by gasoline engine generators \citep{Yuan2008, Zhou2010a}. The observations described in this work took place during the Antarctic winters of 2009 and 2010. 

During the 2009 winter season, observations were carried out by the telescopes equipped with \textit{g}, \cl\ and \textit{r} filters. The remaining telescope, equipped with an \textit{i} filter, failed to return data. The observations used in our analysis span MJD 54935-85 in \textit{g} \& \cl\ and MJD 54955-5143 in \textit{r}. Exposure times varied between $5-20$~s. \citet{Oelkers2015} identified significant operating issues with the telescopes during the 2009 observing season. The primary obstacle for this data reduction was a defocused point-spread function (PSF). The PSF widely varied between regular torus-like shapes and irregular extended torus shapes containing multiple point sources. The \cl\ telescope also suffered from severe intermittent lens frosting which led to the loss of $\sim40\%$ of the data. We chose not to include the \cl\ data in this study because of the relatively small number of data points compared to the other bands and significantly higher dispersion in the resulting light curves \citep{Oelkers2015}. The \textit{r} telescope suffered from computer problems between MJD 54910-55 that resulted in incorrect exposure time information and images in which only part of the CCD appeared to be fully illuminated. Despite the removal of the affected data, $\sim 8\times10^5$ scientifically useful images were acquired over $125$~days \citep{Oelkers2015}.

During the 2010 winter season, observations were only carried out with the telescope equipped with the \textit{i} filter since the other telescopes failed to return data. The observations used in our analysis spanned MJD 55317-5460 with exposures times of $20-40$~s. No major observing complications were identified in previous reductions of this data set or in our analysis.

\section{Image Processing}

All images underwent preliminary data reduction that included bias subtraction, flat fielding, sky background subtraction and an electronic pattern subtraction (in \textit{g} \& \textit{r}) or a fringe pattern subtraction (in \textit{i}). We will briefly describe each technique but refer the reader to the initial data release papers for more detailed descriptions \citep{Wang2013, Oelkers2015}. 

The bias frames used were obtained during instrument testing in China \citep{Zhou2010a} while sky flats were generated by median-combining $>2,700$ frames with high sky level (typically $>10^4$ ADU). After flat fielding, some frames exhibited a low-frequency residual background, likely due to moonlight or aurora. This background was modeled by sampling the sky background every $32\times32$ pixels. Bad and saturated pixels were excluded from each sky sample and a model sky was interpolated between all boxes to make a thin-plate spline \citep{Duchon1976} using the IDL implementation GRID$\_$TPS. This background spline was subtracted from all images.

All images displayed a varying electronic background (in \textit{g} \& \textit{r}) or fringing pattern (in \textit{i}), specially at low sky levels. The electronic patterns were removed from each individual image by masking all sources with $>2.5\sigma$ above the sky level with Gaussian noise, calculating the Fast Fourier Transform, and identifying all peaks with a power greater than 10$^{-3}$. A correction frame was generated using only these peaks and subtracted from the original image. The fringe patterns were removed by masking all stars more than 2$\sigma$ above the sky level and combining all images in sets of 12. The fringe pattern was shown to not change over the 12 images so the final template was created by taking the minimum pixel value at each location among the 12 sets. The resulting template was then subtracted from each science frame. 

\begin{figure}[ht!]
\centering
\includegraphics[width=80mm]{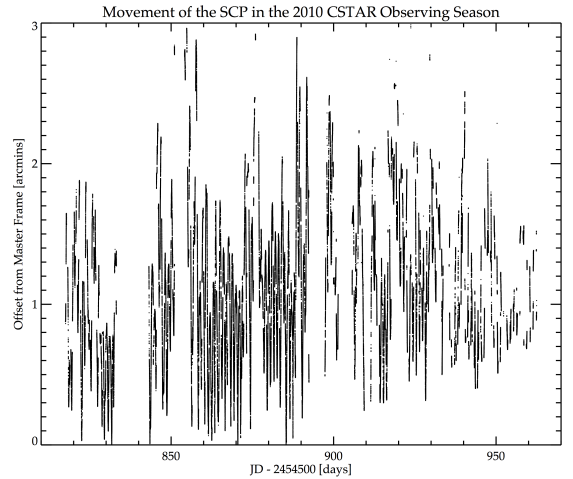}
\caption{The movement of the SCP from the position in our reference frame in \textit{i} as a function of Julian date. There are variations in the movement of the SCP both on a daily level and with a period of 10.5. We hypothesize these periods are due to the changes in the solar elevation, wind, movement of the ice shelf or a combination of all of these factors. \label{fig:scp}}
\end{figure}

Since CSTAR had a fixed pointing towards the SCP, it was necessary to derotate each frame prior to DIA subtraction. We performed initial aperture photometry to all images with DAOPHOT \citep{Stetson1987} and used these star lists with the routines DAOMATCH and DAOMASTER to find the proper cubic transformations between all images. We used these transformations and the IDL routine POLY$\_$WARP to apply a cubic convolution interpolation to align each image. \citet{Oelkers2015} found the SCP to drift slightly throughout the 2009 observing season. We have found the SCP to drift in a similar fashion during the 2010 observing season as shown in Figure~\ref{fig:scp}. The most notable period of the drift is on the sidereal day ($\sim0.997269$~d) with a significant period also found at 10.5~d. We hypothesize these drifts could be due to the change in solar elevation, winds, shifting of the ice shelf or a combination of these factors.

\section{Photometry}

\subsection{Difference Image Analysis}

Difference Image Analysis (DIA) is a photometric technique which uses a convolution kernel to match the seeing conditions between two frames taken of the same star field. A proper kernel solution will subtract any constant flux between frames to zero while leaving true astrophysical changes as correlated residual flux in the differenced frame. DIA has been shown to work well in crowded fields with data from small-aperture telescopes, even with highly abnormal PSF shapes \citep{Pepper2007, Oelkers2015}. Our code is based on the optimal image subtraction routine \textit{ISIS} \citep{AlardLupton} but uses a Dirac-$\delta$ function kernel. The coefficients $c_{\alpha, \beta}(x,y)$ are solved for using the least-squares method by taking stamps around bright, isolated stars. We used a $5\times5$ kernel for all reductions, constant for \textit{i} and first-order for \textit{g} \& \textit{r}, which has been previously shown to work well for CSTAR images \citep{Oelkers2015}. The kernel is defined mathematically below:

\begin{equation}
	K(x,y) =  \sum_{\alpha=-w}^w\sum_{\beta=-w}^w c_{\alpha,\beta}(x,y)K_{\alpha, \beta}(u,v)
\end{equation}

\noindent{where $K_{\alpha, \beta}$ is a combination of $(2w+1)^2$ delta function basis vectors and $K_{0,0}$ is the centered delta function \citep{Miller2008, Oelkers2015}. We allowed $c_{0,0}(x,y)$ to be spatially variable to compensate for imperfect flat field corrections. In the case of $\alpha \neq 0$ and $\beta \neq 0$,}

\begin{equation}
	K_{\alpha, \beta}(u,v) = \delta(u-\alpha,v-\beta) - \delta(u,v) 
\end{equation}

\noindent{while for $\alpha = 0$ and $\beta = 0$,}
\begin{equation}
	K_{0,0}(u,v) = \delta(u,v).
\end{equation}

Clean, precise subtractions require a high signal-to-noise (SNR) reference frame. The 2010 \textit{i} reference frame was created by median combining over $3,500$ images with $>10^4$ stars and a background level below 400~ADU. The 2009 \textit{g} and \textit{r} reference frames are single images devoid of clouds, satellites and lens frosting with low sky background obtained near the end of each observing season. These single reference frames were found to yield substantially better subtractions than co-added images due to the spatially-varying torus-shaped PSFs of the telescopes during that season \citep{Oelkers2015}.

\subsection{Flux Extraction\label{subsec:flux}}

Master star lists were created by first transforming the coordinates of all sources in \citet{Wang2013} to each reference frame. These positions were masked and the DAOPHOT-FIND routine was run to search for any additional stars which might have been missed. We repeated the masking and search steps one additional time to increase completeness at the faint end. We initially removed stars with $<50,000$ data points or which had an rms$>0.75$~mag. The star lists where then matched using the positions from DOAMATCH and DAOMASTER. The final star lists consist of 6,179, 12,801 \& 10,681 stars in \textit{gri} respectively, for a total of 15,496 stars, with 10,094 sources in at least 2 bands. 

We found 292 stars in \textit{g} and 1959 stars in \textit{r} to match to multiple stars in \textit{i} because of the defocused PSF in the 2009 data set. We kept each match and flagged the appropriate stars in the stellar library described in \S~\ref{subsec:lib}. This is shown with the MATCHING flag in Table~\ref{tb:var} with either a 0 (no multi-match) or 1 (multi-match). Similarly some stars in \textit{g} \& \textit{r}, show magnitudes significantly brighter than would be expected for typical \textit{g}-\textit{r} or \textit{r}-\textit{i} colors. This is likely because the aperture magnitudes in the 2009 reference frames likely caused an increase in flux in both bands for some crowded stars. Stars with possible contamination are also flagged in the library with the CROWDING flag with either 0 (normal) or 1 (possible contamination). 

We extracted differential fluxes using aperture photometry with the IDL version of the DAOPHOT package. We set the radius for the photometric aperture at 2.5 and 5 pixels ($37.5$ and $75\arcsec$) for \textit{i} and \textit{g}\&\textit{r}, respectively, with sky annuli spanning $4-6$ and $8-10$ pixels ($1-1.5$ and $2-2.5\arcmin$) in each case. The differential flux was combined with the flux from the reference frame, corrected for exposure time and calibrated following \citet{Wang2011} and \citet{Oelkers2015}. We used the relations from \citet{Wang2013} and \citet{Oelkers2015} to correct for the time drift of the acquisition computer. Finally we removed measurements with large photometric errors and scatter, likely due to cosmic rays or bad subtractions, if these exceeded $+4\sigma$ of a moving median.

\subsection{Detection Frames\label{subsec:detfram}}

DIA provides a unique opportunity to detect variability in a star field before searching through light curves, since correlated residuals on a differenced frame indicate a statistically significant change in flux. A ``detection'' frame can be created by co-adding the absolute values of differenced frames to achieve a higher SNR identification of variable or transient behavior. Each differenced frame was normalized on a per-pixel basis by the square root of the sum of the counts in the science and reference frames before the co-addition. We also masked all pixels within a 5-pixel radius of the position of any point source in the master list.

Due to the nearly-polar location of Dome A, many images were contaminated by satellite tracks which were masked as follows. The FIND routine was used to identify sources in each differenced frame with stellar-like PSFs. These objects were temporarily masked and a line was fit to any remaining pixels with large positive deviations ($>10\sigma$ above the mean background) using the IDL routine ROBUST$\_$LINFIT. If the residuals of the fit had a standard deviation $<3$~pix, a trapezoidal mask was placed along the best-fit line. This process was repeated until no best-fit line was found to account for multiple satellite trails. The temporary masks were then removed and the absolute value of the frame was taken. The final detection frames were made by co-adding all frames obtained within a 24 hour window (typically $>3200$ frames).

\section{Systematic and statistical uncertainties in photometry}

A careful characterization of the sources of systematic and statistical uncertainty is critical prior to claiming the detection of astrophysical transients or stellar variability. Flat-fielding errors, bad subtractions, misalignments and ghost reflections are all examples of sources creating features in light curves which can mimic true signals. We employed several methods, described below, to decrease the impact these systematics played on our data reduction and detection metrics.

\subsection{Bad subtractions due to high attenuation}

DIA uses the least-squares method to find the zero point offset between science frames and a reference frame. In theory, changes due to airmass and the cloud layer should be removed with a proper kernel solution. However, we found that high levels of attenuation due to frosting or clouds ($>1.5$ mag) significantly decrease the SNR of the brightest stars and the number of possible kernel template stars, leading to significant subtraction artifacts. Therefore, we removed any images with $<1300$ stars in the central $512\times512$~pix. This led to a reduction in the dispersion of each light curve by $\sim0.005-0.02$~mag.

\begin{figure}[ht!]
\centering
\includegraphics[width=80mm]{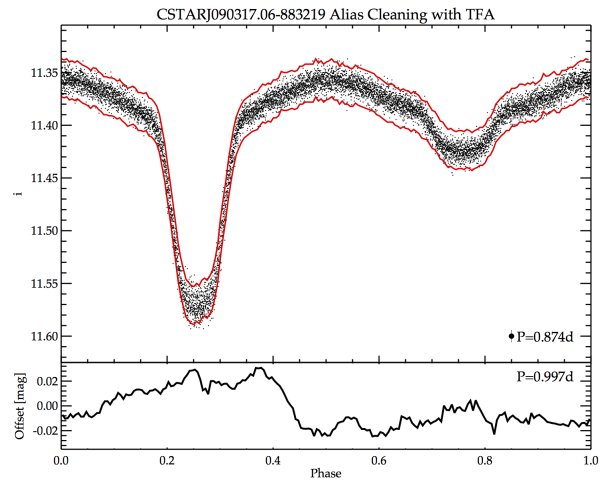}
\caption{\textit{Top}: The phase-folded light curve of a known binary (CSTARJ090317.06-883219) after alias removal. The light curve has been phased and binned into 10,000 data points. The red lines denote the scatter in the light curve prior to the alias removal. The rms has been decreased by $\sim0.017$~mag. Typical photometric error is shown at the bottom right of the panel. \textit{Bottom}: The phase folded aliased signal removed from the light curve. This alias was shown to repeat with a period of the sidereal day. \label{fig:alias}}
\end{figure}

\begin{figure*}[ht!]
\centering
\includegraphics[width=\textwidth]{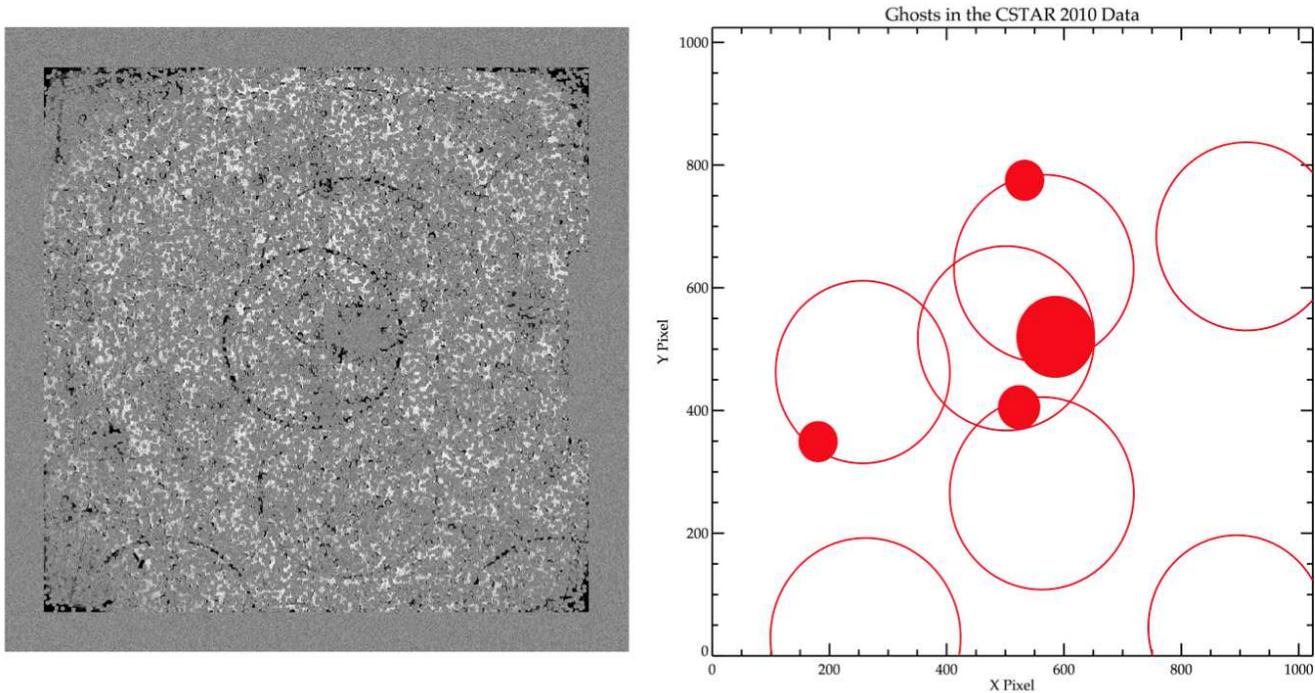}
\caption{The most notable ghosting tracks and saturated stars found in the CSTAR system. \textit{Left}: A 1-day detection frame ($\sim3000$ combined images) showing concentric rings not centered on the SCP likely due to ghost reflections. \textit{Right}: The same ghost rings and the sweeping area of diffraction spikes around bright stars (large red dots) are plotted in x-y space of the 2010 master frame.\label{fig:ghosts}}
\end{figure*}

\begin{figure}[ht!]
\centering
\includegraphics[width=80mm]{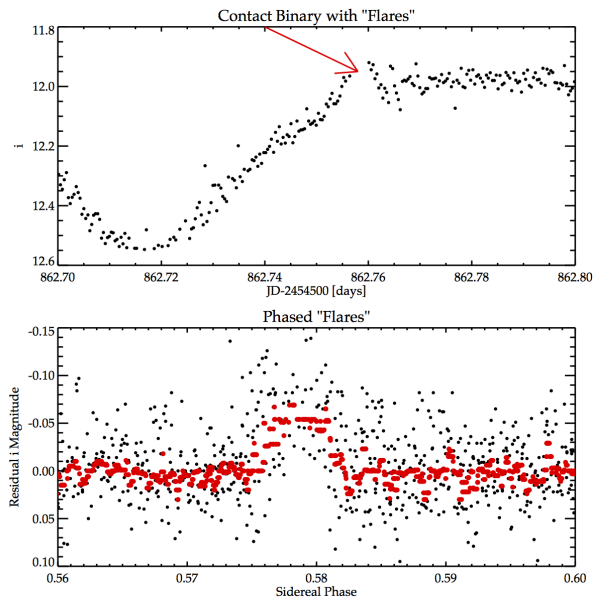}
\caption{\textit{Top}: The light curve of the contact binary CSTARJ084537.33-883343 prior to alias removal. \citet{Qian2014} showed to ``flare" at different times throughout the 2010 observing season. \textit{Bottom}: The same contact binary plotted in phase space for 1 sidereal day after removing the eclipsing variation. All flares quite obviously line up at the same point in phase, signifying the flaring is occurring across the same set of pixels. The red points denote a smoothed 10-minute median of the phased light curve. \label{fig:binary}}
\end{figure}

\subsection{Aliasing}

Despite the continuously dark conditions during the polar night, CSTAR light curves are still significantly affected by aliases with periods close to 1 sidereal day. These are due to the daily circular motion of objects around the FoV. This periodic motion can mimic a true signal when combined with slight flat fielding errors. We found these fluctuations were on the order of $0.01-0.03$~mag in a light curve depending on the stellar magnitude and position on the detector.

Fortunately, since the period of the alias is known ($\sim0.99727$~d) we were able to easily identify and remove these features. We used the Trend-Filtering-Algorithm (hereafter, TFA) in signal reconstruction mode to identify the sidereal signal for each light curve \citep{Kovacs2005, Hartman2008}. The sidereal signal was subtracted from the light curve and TFA was then run iteratively to produce the cleaned product. 150 randomly distributed stars brighter than 11.5~mag with dispersions $<0.1$~mag were used to create the template for TFA. Figure~\ref{fig:alias} shows an example of TFA cleaning a light curve of a known eclipsing binary \textit{without} distorting the astrophysical variation. 

\begin{figure}[ht!]
\centering
\includegraphics[width=75mm]{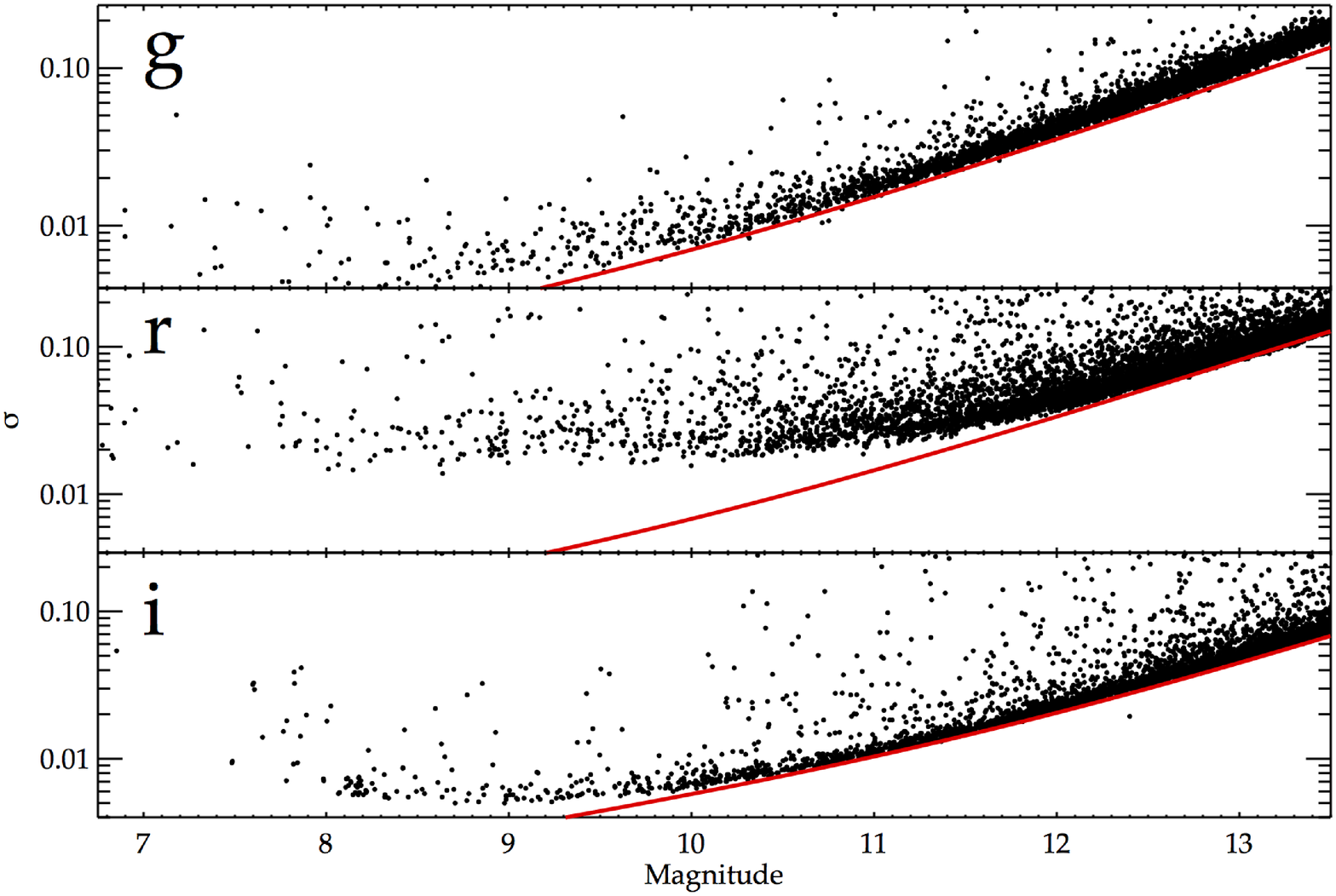}
\caption{The typical daily dispersions for each band with the expected combined/total error from photon noise, sky noise and scintillation plotted as a red line. \label{fig:error}}
\end{figure}

\begin{figure}[h!]
\centering
\includegraphics[width=80mm]{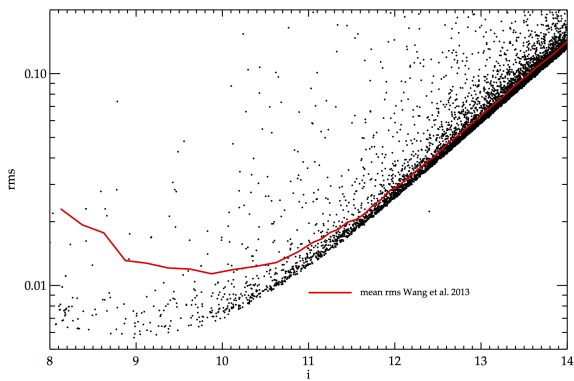}
\caption{A comparison of the rms from this work and the rms of \citep{Wang2013}. The black points mark the rms for stars in the 2010 data set reduced with DIA. The red line is the mean rms of \citet{Wang2013} at each magnitude. This work achieves a lower rms for stars \textit{i}$<11$ by 9~mmag and is consistent with the previous work at fainter magnitudes. \label{fig:comp}}
\end{figure}

\subsection{Ghosts\label{subsec:ghosts}}

The rapid cadence of CSTAR ($20-40$~s) and the long duration of the data stream (nearly 6 uninterrupted months each observing season) allows for the study of stars in an unprecedented fashion. Rare events, such as stellar flares, can be identified and studied in great detail. \citet{Qian2014} showed the modulation of the light curve of a contact binary in the CSTAR field due to the O'Connell effect \citep{OConnell1951} and seemingly aperiodic flaring events throughout the 2010 observing season. Unfortunately, when these flaring events are phased against the sidereal day, they can be traced back to a single $10\times10$ pixel region on the detector. In fact, upon rigorous inspection, $\sim 20\%$ of 2010 CSTAR light curves exhibit similar features which are likely due to internal reflection or ``ghosting'' from bright stars \citep{Meng2013}. Figure~\ref{fig:ghosts} marks the location of the most prominent ghosting features in the CSTAR focal plane and shows the eclipsing binary flares from \citet{Qian2014} as impostor signals. It is very likely that fainter ghosts exist given the large number of stars that exhibit ``flaring events'' which repeat on a sidereal day basis but lie outside of major tracks.

We first identified these ghosting features when creating the daily transient detection frames. As shown in Figure~\ref{fig:ghosts} they appear as circles that are not centered on the SCP. As the ghosts have no source in the reference frame, they appear as seemingly moving objects in the differenced frames. It is likely that these signals are caused by the reflections of bright stars both in and near the CSTAR field, as many of these circles appear to track the reflection pattern of a bright star. These ghosts have been found to make changes in the light curve on the order of $0.2-3$~mag and last $10-45$~min, easily mimicking a flare-like event. Subsequently, any star or transient displaying a flare-like feature had its position checked against these known ghosts and the light curve investigated for similar events occurring at the same point in sidereal phase.

\subsection{Statistical fluctuations}

We modeled the statistical uncertainty as $\sigma^2=I_{N}+A I_{sky}+\sigma^2_a$, where $I_{N}$ and $I_{\rm sky}$ are the photon counts from the object and sky respectively, $A$ is the area of the photometric aperture, and $\sigma_a$ is the expected scintillation limit \citep{Young1967,Hartman2005}. 

We measured the dispersion in each light curve, weighted by the uncertainty in aperture photometry, for a single day of operation (typically $>3000$ frames) and compared these values to our noise model. We find satisfactory agreement with the simple model described above, with dispersions reaching within a factor of 5 of the scintillation limit in \textit{g}\&\textit{i}. We found a larger noise floor of $\sim 1\%$ in \textit{r}, most likely due to the effects of the electronic pattern and the frosting described in \S~2. Figure~\ref{fig:error} shows the typical daily dispersion for each band. 

\section{Search for Variables and Transients}

The combination of robust aliasing removal and use of DIA yielded light curves with lower dispersion than previous analyses of the 2010 CSTAR data, which were based on simpler aperture photometry. We find the rms with DIA to be lower by $\sim9$~mmag for stars with \textit{i}$<11$ as compared to the same stars with aperture photometry as shown in Figure~\ref{fig:comp}. In this section, we discuss the techniques used to analyze this higher-quality set of light curves to search for variability, periodicity and stellar flares. Lastly, we describe the steps taken to search for transients in the ``blank'' areas of the FoV.

\subsection{Continuous Variability\label{subsec:var}}

We employed a combination of 3 variability metrics, following the approach of previous reductions of CSTAR data \citep{Wang2013, Oelkers2015}. We computed the rms and the 90\% span in magnitude for each object (hereinafter, $\Delta_{90}$), as well as the upper 2$\sigma$ envelope of both quantities as a function of magnitude. Light curves above these limits are expected to exhibit true astrophysical variability instead of being dominated by systematic effects. We also employed the Welch-Stetson \textit{J} variability statistic, including the necessary rescaling of DAOPHOT errors \citep{Udalski1994, Stetson1996, Kaluzny1998} as implemented in the VARTOOLS data reduction set \citep{Hartman2008}. This statistic computes the significance of photometric variability between two adjacent data points and is useful to detect variability during short time spans, such as the $5-40$~s sampling of the CSTAR data. We considered objects lying above the $+3\sigma$ envelope for this statistic as variable. 

\begin{figure*}[ht!]
\centering
\includegraphics[width=144.5mm]{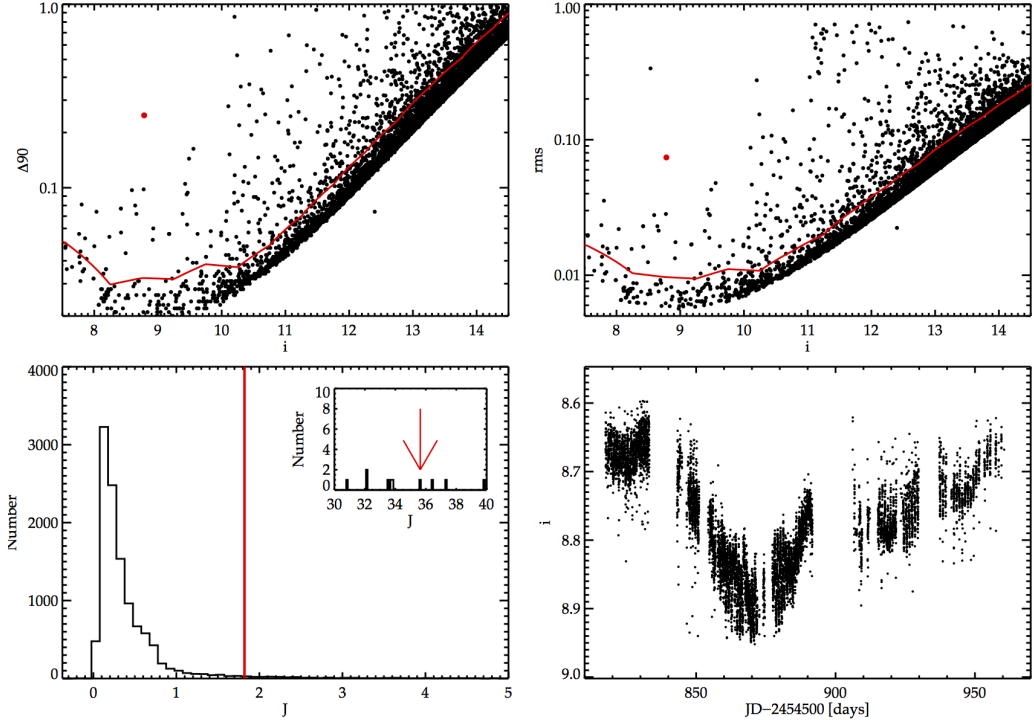}
\caption{Variability tests used to identify variable candidates in the 2009 \& 2010 data sets. Stars lying above the red line in the top panels and to the right of the line in the bottom-left panel are expected to be variable. \textit{Top Left}: $\Delta_{90}$ statistic with the upper 2$\sigma$ quartile plotted as a red line. \textit{Top Right}: rms statistic with the upper 2$\sigma$ quartile plotted as a red line. \textit{Bottom Left}: J Stetson statistic with the upper 3$\sigma$ cut plotted as a red line. \textit{Bottom Right}: The light curve of the variable candidate CSTARJ192723.13-881334 from the 2010 \textit{i} data set. The candidate is shown clearly passing each statistic as a red dot in the top two panels and a red arrow in the bottom-left panel. The light curve is shown in 10-minute bins with the size of each data point being the size of the typical photometric error.\label{fig:stat}}
\end{figure*}

We considered a star to be variable if the star passed all 3 of the above tests in either \textit{g}, \textit{r} or \textit{i}. A star was removed from the variable sample if it was within 3.75 (7.5) pixels of a star 2 magnitudes brighter in \textit{i}(\textit{g\&r}) or the star's \textit{primary} LS period was an aliased period with SNR greater than 1$\sigma$ of the mean SNR for a given band. We also removed the $10\%$ of stars with the least amount of data. A star was returned to the periodic sample if it was later found to have a significant LS period which was \textit{not} an alias and passed the other 2 cuts described above. Figure~\ref{fig:stat} shows these techniques recovering the variable candidate CSTARJ192723.13-881334. 

\begin{figure*}[ht!]
\centering
\includegraphics[width=144.5mm]{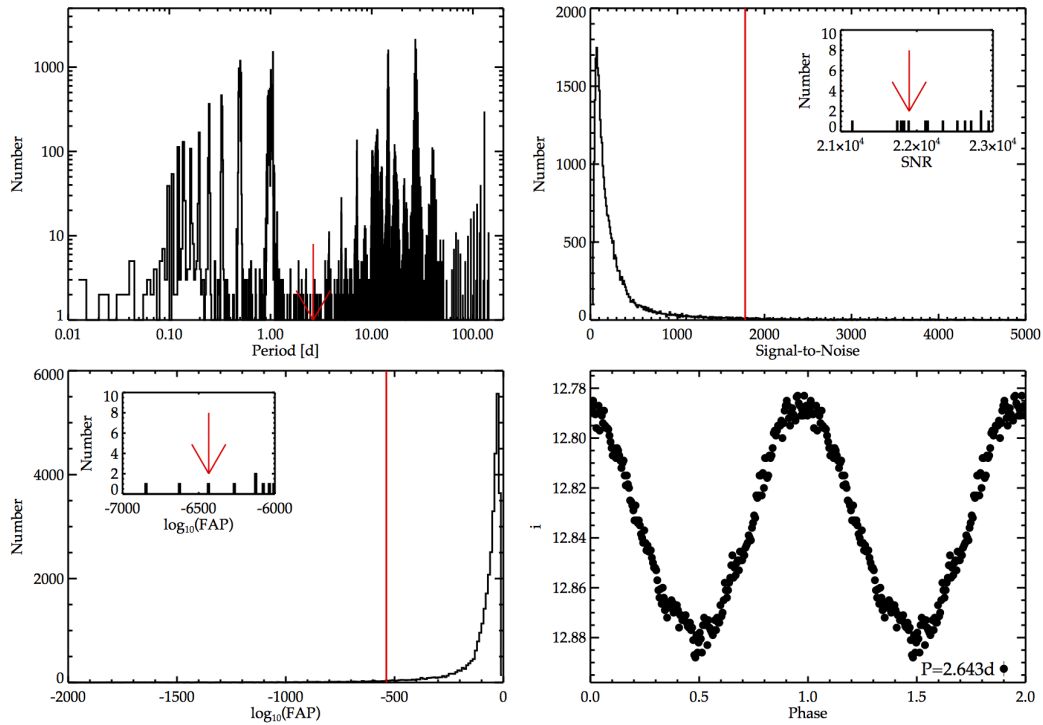}
\caption{The periodicity tests used to identify periodic candidates in the 2009 \& 2010 data sets. \textit{Top Left}: the number of ``variable'' stars with similar periods, indicative of aliasing. The passing candidate is shown with a red arrow. Notice the period is not found on or near a large distribution of other periods; \textit{Top Right}: the $+3\sigma$ cut (red line) on the signal-to-noise ratio. The passing candidate is shown with an arrow; \textit{Bottom Left}: the $+3\sigma$ cut (red line) on the false alarm probability. The passing candidate's log$_{10}$(FAP) is shown with an arrow; \textit{Bottom Right}: The light curve of a periodic variable star candidate CSTARJ071218.39-875116. The light curve has been phase folded on the recovered period of 2.64~d, binned into 200 data points and plotted twice for clarity. The typical error is shown at the bottom right of the panel. \label{fig:perd}}
\end{figure*}

\clearpage
\subsection{Classical Periodicity}

We searched each light curve for periodic signals using a Lomb-Scargle periodogram \citep[LS]{Lomb, Scargle} as implemented in VARTOOLS \citep{Hartman2008}. We computed the 3 highest SNR periods of each star between 0.01~d and the total number of days observed in each band. Each light curve was whitened against the highest SNR period before searching for the next. We checked each signal against know aliases and removed spurious signals. We also applied $+3\sigma$ cuts false alarm probability (log$_{10}$(FAP)) and SNR. The FAP provides an estimate on the likelihood of a true periodic signal by comparing the SNR of a specific signal to the cumulative distribution of all SNRs. Figure~\ref{fig:perd} shows the technique recovering the period for the candidate  CSTARJ071218.39-875116. We removed stars from the periodic sample using the same cuts described in \S~\ref{subsec:var}.

We also searched for transit-like events using the Box-Least-Squares algorithm \citep[BLS]{Kovacs2002}. We pre-whitened each light curve against the most significant LS period and its 10(9) (sub)harmonics. The duration of the transit was allowed to range in phase between 0.01 and 0.1 of the primary period with 10,000 trial periods and 200 phase bins. The search was conducted between 0.1 and $1/3$ of the total days in the observing season in order to ensure we observed \textit{at least} 3 transit events. We required each eclipse period to be unique and have a unique ephemeris time to avoid contamination by aliased periods. We also required the ratio of $\chi^{2}/\chi^{2}_{-}>1.0$ and the signal-to-pink noise to be $>5$. The $\chi^{2}/\chi^{2}_{-}$ statistic shows how well a transit model is fit compared to a transit model fit to the inverted light curve \citep{Burke2006}. Light curves with $\chi^{2}/\chi^{2}_{-}<<1$ are ill suited for transit searches. Each periodic variable passing the BLS cuts was visually inspected to ensure the routine was not fitting the noise.

\subsection{Stellar Flares\label{subsec:flr}}

We searched for flare events using the IDL function GAUSSFIT with a 6-term solution to allow for symmetric and asymmetric flare detection. Similar to the BLS search, we pre-whitened all light curves against the most significant LS period and its 10(9) (sub)harmonics. Each whitened light curve was broken into 0.25~d bins with at least 50 data points per bin prior to the fit. Any best-fit gaussian with $0.8<\chi^2_{\nu}<1.2$ and a flare amplitude greater than the rms of the light curve passed the first significance cut. 

All passing events were phased on the sidereal day to check against ``ghosting'' signals. Any recurrent event in sidereal phase was flagged as a spurious ``ghost" detection. This procedure was repeated, phasing the events on the whitened LS period to identify events which may have been artifacts of the whitening process. Similarly, the MJD of each flare was checked against all other flare event timings to rule out events which were caused by global artifacts, such as misalignments or bad subtractions. The phase of each flare event was also visually inspected to confirm there were no other noticeable flares, indicative of ghost events and the previous and next sidereal day were examined no similar variation occurred. Additionally, any star with a candidate flare in \textit{g} or \textit{r} observed between MJD 54955-985, when the observations overlapped, had its light curve inspected in the alternate band. If the flare did not pass the cuts mentioned above then the flare was removed from the candidate list.

Any flare timing within 5~min of a flare in \textit{another} star was flagged and removed. Each flaring star's position was required to be more than 5~pixels from a known ghosting track or bleed trail from a saturated star. Candidates were further constrained to have $0.95<\chi^2_{\nu}<1.05$ to remove candidates which were fitting the noise of a light curve instead of flare-like variation. Lastly, we removed stars from our flaring sample which did not pass the cuts for proximity or aliasing mentioned in  \S~\ref{subsec:var}.

We attempted to quantify the possible ghost contamination in our sample because of the large number of light curves showing ghosting events. We estimated this contamination by injecting fake flares of varying amplitude, length and phase into simulated light curves with varying noise. These contaminated light curves were then run through our selection process. We found on average $12\%$ of the total flares recovered were ghost contaminants with lengths $<45$~min. We use this contamination rate to correct our flaring fraction.

To quantify the flare rate and make comparisons to the previously mentioned studies we needed to select the stars which were the most likely to be K/M dwarfs. We identified the stars in our data set using the 2MASS catalogue \citep{Skrutskie2006} to provide \textit{JHK} magnitudes. We combined the \textit{J-H} vs.~\textit{H-K} color-color diagram with the stellar locus for K5V-M9V provided by \citet{Pecaut2013}. We selected stars with 2MASS photometric errors $\sigma < 0.2$~mag and within $\pm1\sigma$ of the \textit{J-H} v. \textit{H-K} loci as the most likely dwarf candidate members. To estimate contamination by background giants we queried the TriLegal model \citep{Girardi2012} for the Galaxy and applied the same cuts. We estimate our contamination to be $<1\%$ at each spectral type. We estimate the Galactic reddening vector with the relations from \citet{Fitzpatrick1999} and find extinction would preferentially scatter early type dwarfs into our selection sample. However, since SCP E(B-V) at the SCP will be $\sim0.16$~mag \citep{Schlafly2011}, the expected color excess in the 2MASS bands is E(J-H)$<0.05$~mag and E(H-K) $< 0.03$~mag.  We expect these effects to cause minimal contamination from early type stars.

\subsection{Transient Events}

The detection frames, described in \S~\ref{subsec:detfram}, were inspected for correlated residuals with stellar-like PSFs in the ``blank'' areas of the master frame. Recall that all point sources detected in the master frame were masked in these detection frames; therefore this search was specifically aimed at identifying transients arising from objects normally lying below the limiting magnitude of CSTAR. 7$\times$7 pixel stamps centered on each transient candidate were extracted and retained if they exhibited a $>+5\sigma$ variation above the sky background. If a transient candidate occurred in \textit{g} \& \textit{r} between MJD 54955-985 its position and timing were checked in the alternate band to aid in confirmation. Any transient without a counterpart was removed from our sample.

Fluxes were then extracted from all differenced frames (as described in \S~\ref{subsec:flux}) for two reasons. The first was to check for \textit{bona-fide} variation of the transient light curve, which might have been missed by the metric above. The second was to give a robust sample of possible aliasing flares described in \S~\ref{subsec:ghosts}. Since the majority of each transient light curve was simply the sky background, variations due to moonlight or twilight had to be removed by subtracting the median sky value of the image from each transient light curve. 
\begin{deluxetable*}{cccccccccccccc}[b!]
\tabletypesize{\tiny}
\tablewidth{0pt}
\tablecaption{Difference Image Stellar Library* \label{tb:var}}
\tablehead{\colhead{ID} &  \colhead{R.A.} & \colhead{Dec.} & \colhead{\textit{g}} & \colhead{\textit{r}} & \colhead{\textit{i}} & \multicolumn{3}{c}{Metrics} & \colhead{Type} & \multicolumn{2}{c}{Period [d]} &  \colhead{Crd} & \colhead{Mch} \\
& & & & &  & \colhead{V} & \colhead{P} & \colhead{T} & & \colhead{LS} & \colhead{BLS} & \\}
\startdata
CSTARJ142823.83-883839&14:28:23.83&-88:38:39&12.589&12.624&12.115&4&4&0& RR&  0.646542&       \nd&0&0\\
CSTARJ064036.43-881422&06:40:36.43&-88:14:22&12.095&11.960&11.727&4&4&0& EB&  0.438630&       \nd&0&0\\
CSTARJ092839.06-882923&09:28:39.06&-88:29:23&12.547&12.065&11.929&4&4&0& MP&  0.621802&       \nd&0&1\\
CSTARJ092823.53-882925&09:28:23.53&-88:29:25&12.227&12.065&11.733&4&4&0& MP&  0.621805&       \nd&0&1\\
CSTARJ084537.33-883343&08:45:37.33&-88:33:43&13.175&12.570&11.996&4&3&2& EB&  0.267136&  0.267063&0&0\\
CSTARJ100346.70-884425&10:03:46.70&-88:44:25&12.508&10.536&12.853&3&0&0& IR&       \nd&       \nd&1&1\\
CSTARJ030027.40-880305&03:00:27.40&-88:03:05& 0.000&12.419&10.104&3&0&0& LT&       \nd&       \nd&0&0\\
CSTARJ024229.86-880426&02:42:29.86&-88:04:26&12.238&11.220& 9.466&3&0&0& LT&       \nd&       \nd&0&0\\
CSTARJ111627.63-883533&11:16:27.63&-88:35:33&12.014&10.984& 9.386&3&0&0& LT&       \nd&       \nd&0&0\\
CSTARJ100122.86-884438&10:01:22.86&-88:44:38&11.170&10.536&10.079&3&2&0& LT& 43.779884&       \nd&0&1\\
... & ... & ... & ... & ... & ... & ... & ... & ... & ... & ... & ...\\
\enddata
\tablecomments{*:The full table is available for download with the stellar library. The metric flags (V - variability, P - periodicity, T - transit/BLS) mark the scores mentioned in \S~\ref{subsec:var}. The crowding flag (Crd) marks stars which may be showing brighter magnitudes in a band due to crowding. The matching flag (Mch) marks stars which may match to more than 1 star in different bands due to blending.}
\end{deluxetable*}

\begin{figure*}[ht!]
\centering
\includegraphics[width=120mm]{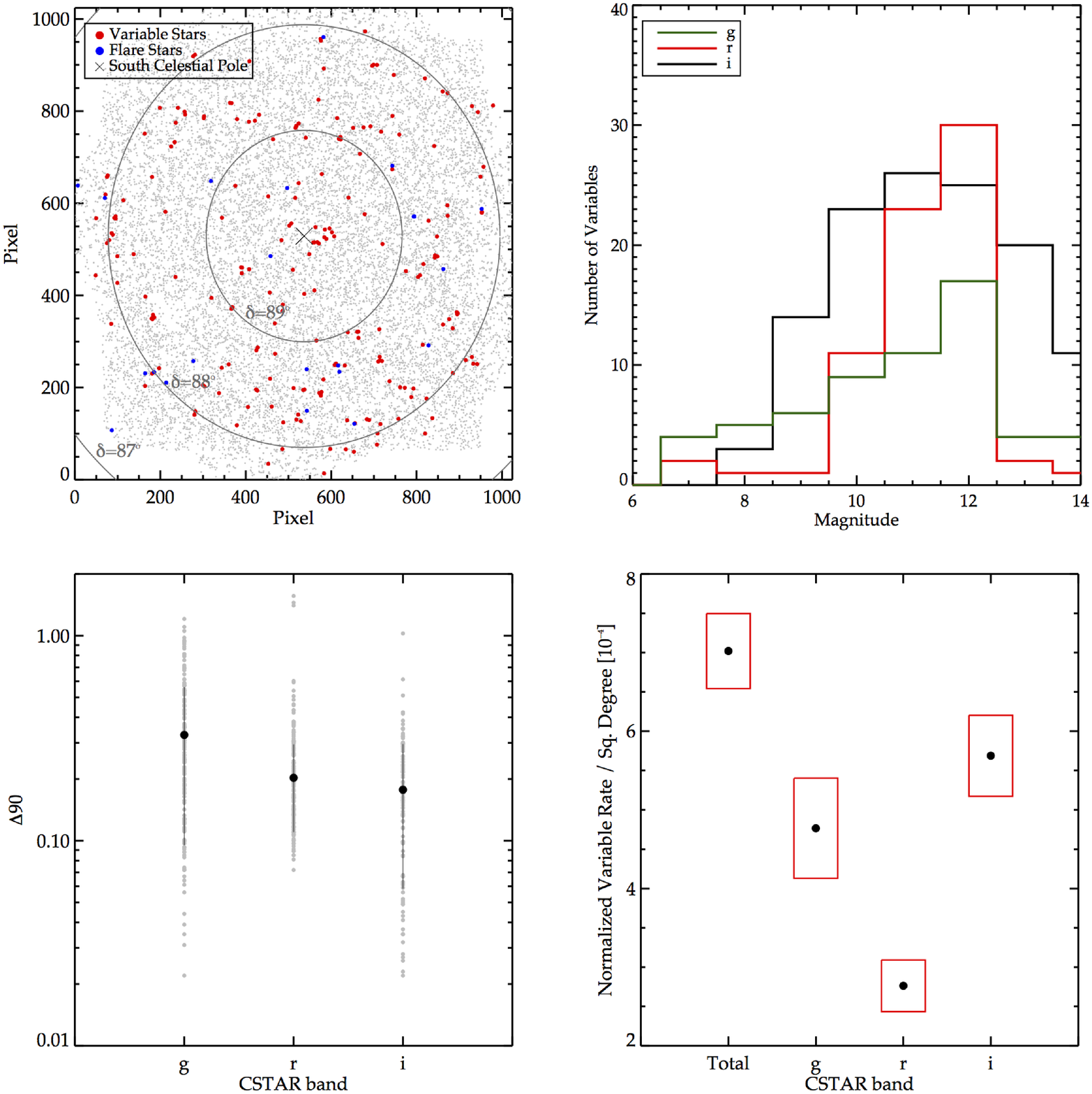}
\caption{\textit{Top Left}: Positions (in CSTAR detector coordinates) of stars in our library (grey dots). Catalogued continuously variable stars are shown with red dots; flaring stars are shown with blue dots; the cross marks the SCP. The points appear to be randomly distributed across the detector. \textit{Top Right}: The number of variable stars identified as a function of magnitude for a given band. \textit{Bottom Left}: The $\Delta_{90}$ statistic for all identified variable stars in a given band. Variability appears to be larger in \textit{g}. \textit{Bottom Right}: The normalized variable star rate for a given square degree of the sky. All errors are based on Poisson statistics and agree with previous reductions of the CSTAR data sets.\label{fig:varrate}}
\end{figure*}

Each transient candidate light curve was checked against known aliasing features as follows. The light curve was divided into segments spanning 0.01 sidereal days and the mean magnitude of each fragment was compared to $\bar{m}_\phi$, defined as the mean magnitude of all other sections of the light curve spanning the same fractional sidereal day during the rest of the season.  The variation was considered \textit{bona-fide} if it contained at least 10 data points and lay $>+2\sigma$ above $\bar{m}_\phi$. The timing of each transient passing these cuts was further checked against the timing of \textit{all} others. If an event was found to coincide in time with another candidate, both were discarded as spurious.

\section{Results and Conclusions}

\noindent{\it Classic variability and/or periodicity:\label{subsec:lib}} Previous studies of CSTAR data \citep{Wang2011, Wang2013, Oelkers2015, Yang2015} have generated lists of variables by applying a binary classification (i.e., an object is either variable or not, based on a set of criteria). In this work, we present the likelihood of variability and/or periodicity for every object, computed as follows. Stars meeting the variability criteria in \textit{g} or \textit{r} received one point per band, while two points were awarded for \textit{i} because those images were in focus and well sampled. Similarly, stars exhibiting a significant periodicity received 2 points in \textit{i} and 1 in \textit{g} \& \textit{r}. We found 48 objects to have a variability or periodicity (LS or BLS) score of 3 or more, signifying a high likelihood of variability. Table~\ref{tb:var} is an example list of all stars in our sample and their resulting scores which will be included with the stellar library. If the star had a variability or periodicity score of 3 or more, its type was estimated in Table~\ref{tb:var}.

Numerous reductions of the CSTAR data sets have identified many new and intriguing variable stars. The unprecedented cadence of the telescope over a 6 month period allows for a statistical analysis of the number of variable stars which could be visible in a given FoV. Figure~\ref{fig:varrate} shows all of the variable stars in our field as well as the flaring stars described below.  We determined our variable star rate by looking at the total number of stars passing the variability tests (rms, $\Delta_{90}$, \textit{J}) in at least 1 band. We excluded the periodic variables (both LS and BLS) from this analysis unless their variations were large enough to pass our tests for classical variability. We find the majority of our recovered variable stars are mag$\sim12$ in all bands with variables in \textit{g} showing the largest magnitude variation. The large number of variables at $12$~mag is likely due to the fact that $\sim40$ of the stars observed were at this magnitude. Finally we determined a normalized variable rate of $7.0\pm0.5\times10^{-4}$ variable stars per sq. degree across all bands, $4.5\pm0.6\times10^{-4}$ for \textit{g}, $2.8\pm0.3\times10^{-4}$ for \textit{r} and $5.7\pm0.5\times10^{-4}$ for \textit{i}. These rates are consistent with previous studies of the CSTAR field.

\begin{figure*}[ht!]
\centering
\includegraphics[width=\textwidth]{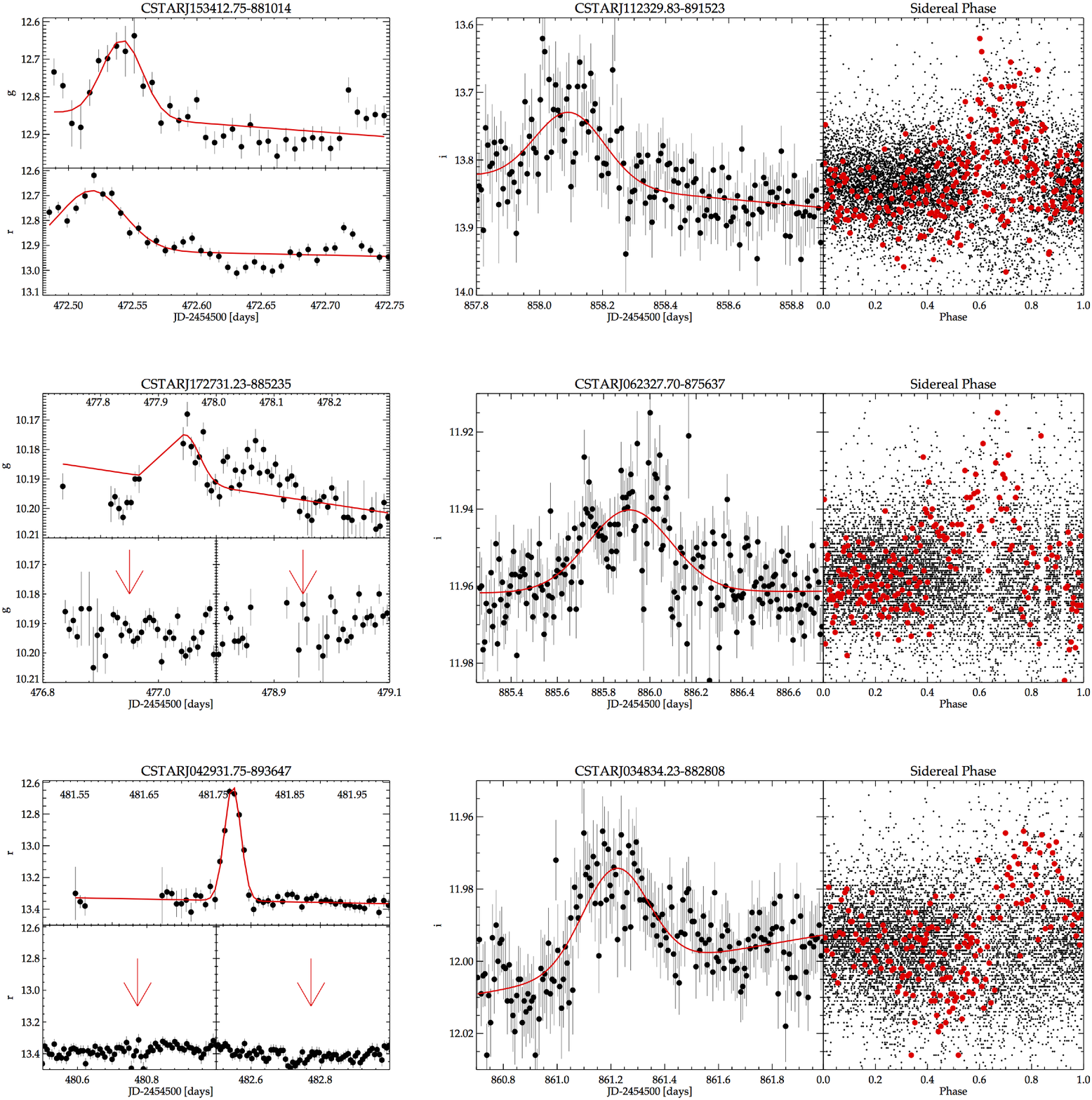}
\caption{Light curves of 9 flaring events, with various lengths and amplitudes, all flagged as genuine in our search. Each panel shows a particular technique to remove ghost contamination. \textit{Top Left}: The flare appears in both \textit{g} \& \textit{r}. \textit{Bottom and Middle Left}: The flare does not occur in the previous or next sidereal day, suggesting a genuine event. \textit{Right}: The flares do not appear to coincide with any noticeable features when phased on the sidereal day. Red points in the phase-folded light curve denote all photometric points visible in the left panel. All light curves have been binned in 10-minute intervals. \label{fig:flares}}
\end{figure*}

\begin{figure*}[ht!]
\centering
\includegraphics[width=150mm]{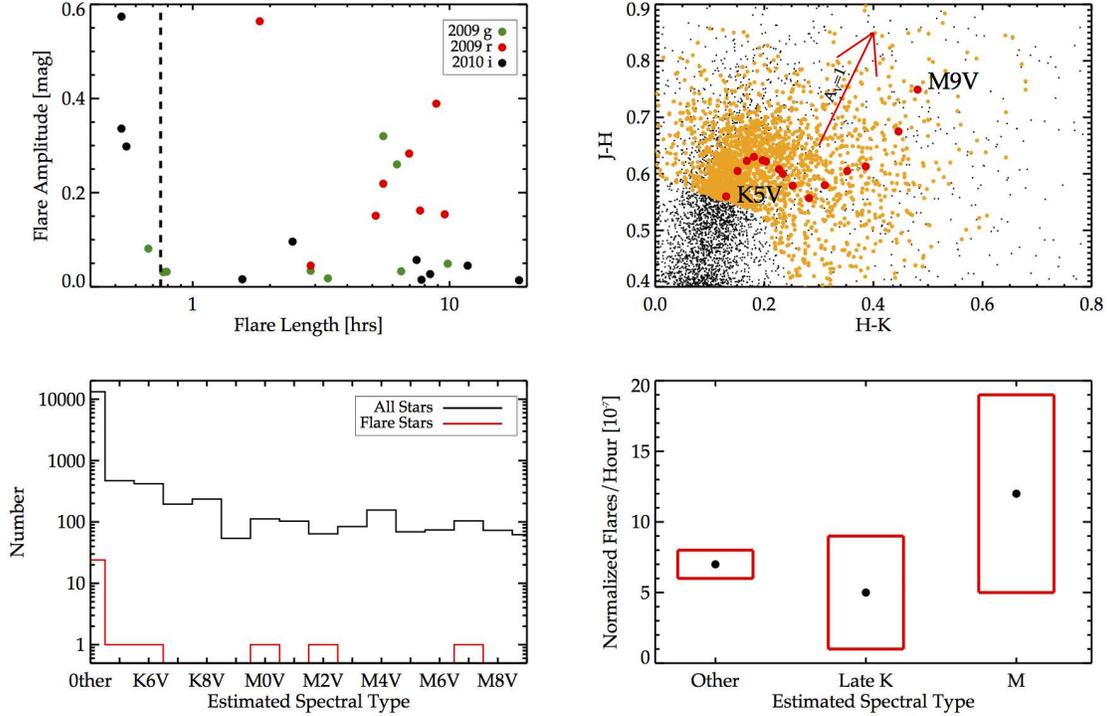}
\caption{\textit{Top Left}: Flare timing vs. flare amplitude for all events flagged as genuine. The dashed line shows the maximum length of ghosting events. Any event to the right of this line is likely to \textit{not} be a contaminating ghost. \textit{Top Right}: Selection of dwarf stars using 2MASS J-H vs H-K colors and the stellar locus of \citet{Pecaut2013}. Stars with 2MASS colors falling within $1\sigma$ of a stellar locus and with 2MASS photometric error $<0.2$~mag were selected as dwarf candidates. The reddening vector shown is based on the extinction law from \citet{Fitzpatrick1999}. \textit{Bottom Left}: Histograms for the total number of late K and M dwarfs in the CSTAR data set. The red histogram shows flaring stars in our sample. \textit{Bottom Right}: Normalized flare rates derived from our observations and errors are based on Poisson statistics. All results are consistent given their uncertainties. \label{fig:rate}}
\end{figure*}

\begin{figure}[ht!]
\centering
\includegraphics[width=80mm]{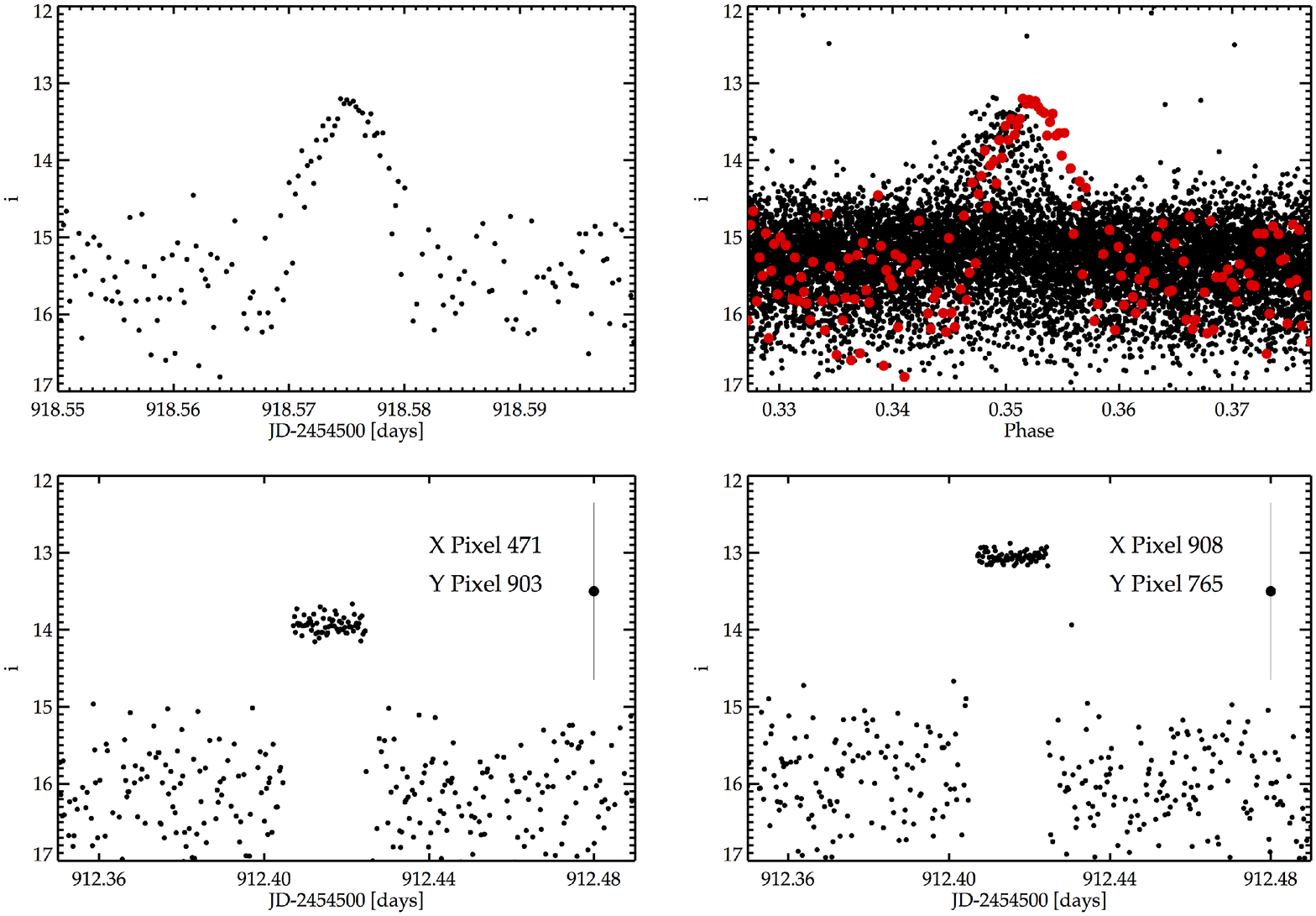}
\caption{\textit{Top Left}: The light curve of a transient candidate in \textit{i} identified in a detection frame with a $\sim 2$~mag variation that lasted for 0.01~d. \textit{Top Right}: The same transient from the left phased on the sidereal day. The red data points mark the data points from the top-left panel. A similar variation is seen near the same sidereal phase and therefore excludes the transient as a real candidate. \textit{Bottom Panels}: Two separate transient candidates showing a sudden increase in flux that remains constant for the duration of the event. Both candidates are ruled out because of the simultaneous occurrence of the events. The typical error and (x,y) location is shown on the right of each panel.\label{fig:bdtran}}
\end{figure}

\noindent{\it Flare events:} We identified 10, 10 and 9 flare events in \textit{i, g} \& \textit{r}, respectively leading to a total of 29 flares throughout the nearly $3000$ combined hours of observations between 2009 and 2010. This leads to a total flaring rate for the entire CSTAR field of $7\pm1\times10^{-7}$ flares/hr. Details for each flare event are shown in Table~\ref{tb:tran} and Figure~\ref{fig:flares} shows light curves of 9 events of varying amplitude and length. Of the stars which could have been visible in both \textit{g} \& \textit{r} we found only 1 star to flare in both bands. The remaining objects either did not have a counterpart in the other band, were outside of the counterpart band's observing window or experienced a data drop out at the time of the flare. 

The normalized flare rates for the searched spectral types, K5V-M9V, are  $5\pm4\times10^{-7}$ flares/hr (late K) and $1\pm1\times10^{-6}$  flares/hr (M) as shown in Fig~\ref{fig:rate}. All other stars in our sample were shown to flare at a rate of $7\pm1\times10^{-7}$ flares/hr. We found 6 stars in this spectral range to flare which is consistent with our expectations of $1-4$ flaring K/M dwarfs from the previously defined flaring fraction. These rates are in contention with previous studies of the flare rates for these spectral types, $\approx 4\times10^{-4}-10^{-1}$ flares/hr \citep{Davenport2012, Hawley2014}. These rates were shown to be highly dependent on the activity level of the star.

We hypothesize our flare rates are lower because of a combination of factors. First, the relative age of stars in the halo is typically older than that of stars in the disk \citep{Jofre2011}. \citet{Kowalski2009} showed the flare rate was highly dependent on galactic latitude. As the CSTAR field is centered at ($l\approx303$, $b\approx-27^{\circ}$) it is dominated by halo stars \citep{Wang2013, Oelkers2015}. Studies of stellar rotation and activity relations for diverse stellar ages have shown older stars decrease their magnetic activity as they age \citep{Garcia2014}. Therefore an older, magnetically inactive halo population would flare at a lower rate than a more diversely aged population as is found in the disk (i.e. stars in the Kepler field \citep{Hawley2014}). Similarly, \citet{Hawley2014} showed inactive M-dwarfs could have flare rates lower than active M-dwarfs by 2-3 orders of magnitude.  

Finally, a major contributing factor to our lower flare rate is that we are biased against detecting \textit{all} flares due to the rampant ghosting events in the CSTAR data sets. Many of the flares contributing to the flare rates in previous work were found with significantly more precise photometry, from the Kepler space telescope, and had durations $<1$~hr, a timescale identical to ghost reflections \citep{Hawley2014, Davenport2014, Lurie2015}. We corrected our flaring rate for ghost contamination by subtracting our expected contamination rate for flares with timescales less than 45 minutes as determined in \S~\ref{subsec:flr} but because we made many cuts on simultaneous events, sidereal phase timing and flare duration we likely removed \textit{bona-fide} flares from our sample. If the telescope had returned more simultaneous, multi-band data during the 2009 or 2010 seasons we would be able to better constrain, identify and categorize more flare candidates.

\noindent{\it Transient events:} We identified 331, 53 and 15 candidate transient events in \textit{i, g} and \textit{r}, respectively. However, throughout our analysis it became quite clear many systematic effects can mimic transients, specially at low flux levels. After studying the timing of each event as well as its duration and amplitude, it became clear that \textit{none} of the candidates could be distinguished from known detector systematics. Many identified transients either showed no variation in their light curve at the expected time or exhibited signals which mimicked those found in \S~\ref{subsec:ghosts} at nearly the same fractional sidereal day ($\pm$ 0.01). Other transients occurred only once during the season, but were found to occur at the same time as other events elsewhere in the focal plane. Similarly, transients which should occur in both \textit{g} \& \textit{r} where unrecoverable in the counterpart band. Figure~\ref{fig:bdtran} shows examples of such impostors. While no events were identified, our null result is consistent with the expectation of the type 1a supernovae occurrence rate being less than $2\%$ based on CSTAR's limiting magnitude and FoV.

We find that CSTAR's original design may not have been well suited for blind transient searches due to its large pixel scale ($\sim15\arcsec$) and lack of a tracking mechanism. The large pixel scale forced many sources into the same pixel and also exacerbated the difficulty of locating the source of the transient event in a possible 1.25\arcmin radius. The large pixel scale also created a shallow limiting magnitude (13.5 in \textit{g}\&\textit{r} and 14.5 in \textit{i}) which greatly reduced the telescope's ability to detect Galactic and extragalactic transients. Furthermore, the lack of a tracking mechanism increased the number of sources effected by ghosting events. We find these events to be so common and wide spread that if a short-duration transient event were to be detected with no other known counterpart it is more likely the event is caused by an asymmetric reflection than truly being astrophysical in origin. 

In contrast, successful transient searches, such as the Palomar Transient Factory and the Dark Energy Survey, use significantly larger telescopes with finer pixel scales \citep{Law2009, Kessler2015}. The ASAS-SN survey, which uses a similar aperture telescope, employs a mount capable of tracking and has a fainter limiting magnitude of 17 in V \citep{Shappee2014}. Nevertheless, CSTAR was quite useful for delivering high SNR light curves of bright variable stars and for detecting periodic signals such as planetary transits and stellar eclipses. We suggest decreasing the pixel scale and adding a tracking mechanism for any future plan of returning the telescope to Dome-A. However, given the difficulty of installing and operating a robotic telescope from Dome A, we consider the telescope to be highly successful because of the quality of the photometry and numerous scientific studies resulting from the 3 years of operation \citep{Zou2010, Zhou2010, Wang2011, Wang2013, WangS2014, Qian2014, Oelkers2015, WangS2015, Yang2015}.

\section{Summary}

We have presented an in-depth search for transients, stellar flares and variables in the 2009 and 2010 CSTAR observations, using difference-imaging analysis. The improved photometry delivered by this technique enabled us to search for stellar flares and to better characterize classical (and in some cases periodic) variables. We identified 29 flaring events implying a flare rate of $7\pm1\times10^{-7}$ flare/hr for the entire CSTAR field, $5\pm4\times10^{-7}$ for Late K and $1\pm1\times10^{-6}$ flare/hour for M dwarfs in the halo. Our flare rate for all other stars in our sample is $7\pm1\times10^{-7}$ flare/hr. We also identify 48 of 15,496 stars with highly significant continuous variability or periodicity, which is consistent with previous reductions of variable stars in the CSTAR field. We determine a total variability rate of $7.0\pm0.5\times10^{-4}$ variable stars per sq. degree which could be useful in planning future high-cadence searches for variable stars. 

The generation of ``detection frames'' for the transient search enabled us to identify a large number of systematic effects, some of which have resulted in the erroneous claim of astrophysical variability in previous analyses of these data sets. We have detailed these systematic effects to help future users of the photometric data products identify possible contaminants. Small-aperture telescopes can be very beneficial in blind transient searches if they have fine pixel scales and tracking capabilities. Additionally, they are very well suited for not long-term studies of bright variables and can provide unique information on light curve modulations or apsidal motion.

All light curve data will be available through the Chinese Virtual Observatory (http://casdc.china- vo.org/data/cstar) and the difference-image-analysis code developed for this analysis is freely available upon request to the corresponding author (RJO).

\acknowledgments
RJO \& LMM acknowledge support from the George P. and Cynthia Woods Mitchell Institute for Fundamental Physics and Astronomy and the Mitchell-Heep-Munnerlyn Endowed Career Enhancement Professorship in Physics or Astronomy. We thank the anonymous referee for their valuable comments, which greatly improved the quality of this paper.

The support of the Australian Research Council and the Australian Antarctic Division is acknowledged. Iridium communications were provided by the US National Science Foundation and the United States Antarctic Program.

The support of Chinese Polar Environment Comprehensive Investigation \& Assessment Program is acknowledged.

\bibliographystyle{apj}
\bibliography{references}

\begin{thebibliography}{}
\expandafter\ifx\csname natexlab\endcsname\relax\def\natexlab#1{#1}\fi

\bibitem[{{Alard} \& {Lupton}(1998)}]{AlardLupton}
{Alard}, C., \& {Lupton}, R.~H. 1998, \apj, 503, 325

\bibitem[{{Bakos} {et~al.}(2002){Bakos}, {L{\'a}z{\'a}r}, {Papp}, {S{\'a}ri},
  \& {Green}}]{Bakos2002}
{Bakos}, G.~{\'A}., {L{\'a}z{\'a}r}, J., {Papp}, I., {S{\'a}ri}, P., \&
  {Green}, E.~M. 2002, \pasp, 114, 974

\bibitem[{{Borucki} {et~al.}(2010){Borucki}, {Koch}, {Basri}, {Batalha},
  {Brown}, {Caldwell}, {Caldwell}, {Christensen-Dalsgaard}, {Cochran},
  {DeVore}, {Dunham}, {Dupree}, {Gautier}, {Geary}, {Gilliland}, {Gould},
  {Howell}, {Jenkins}, {Kondo}, {Latham}, {Marcy}, {Meibom}, {Kjeldsen},
  {Lissauer}, {Monet}, {Morrison}, {Sasselov}, {Tarter}, {Boss}, {Brownlee},
  {Owen}, {Buzasi}, {Charbonneau}, {Doyle}, {Fortney}, {Ford}, {Holman},
  {Seager}, {Steffen}, {Welsh}, {Rowe}, {Anderson}, {Buchhave}, {Ciardi},
  {Walkowicz}, {Sherry}, {Horch}, {Isaacson}, {Everett}, {Fischer}, {Torres},
  {Johnson}, {Endl}, {MacQueen}, {Bryson}, {Dotson}, {Haas}, {Kolodziejczak},
  {Van Cleve}, {Chandrasekaran}, {Twicken}, {Quintana}, {Clarke}, {Allen},
  {Li}, {Wu}, {Tenenbaum}, {Verner}, {Bruhweiler}, {Barnes}, \&
  {Prsa}}]{Borucki2010}
{Borucki}, W.~J., {Koch}, D., {Basri}, G., {et~al.} 2010, Science, 327, 977

\bibitem[{{Burke} {et~al.}(2006){Burke}, {Gaudi}, {DePoy}, \&
  {Pogge}}]{Burke2006}
{Burke}, C.~J., {Gaudi}, B.~S., {DePoy}, D.~L., \& {Pogge}, R.~W. 2006, \aj,
  132, 210

\bibitem[{{Davenport} {et~al.}(2012){Davenport}, {Becker}, {Kowalski},
  {Hawley}, {Schmidt}, {Hilton}, {Sesar}, \& {Cutri}}]{Davenport2012}
{Davenport}, J.~R.~A., {Becker}, A.~C., {Kowalski}, A.~F., {et~al.} 2012, \apj,
  748, 58

\bibitem[{{Davenport} {et~al.}(2014){Davenport}, {Hawley}, {Hebb},
  {Wisniewski}, {Kowalski}, {Johnson}, {Malatesta}, {Peraza}, {Keil},
  {Silverberg}, {Jansen}, {Scheffler}, {Berdis}, {Larsen}, \&
  {Hilton}}]{Davenport2014}
{Davenport}, J.~R.~A., {Hawley}, S.~L., {Hebb}, L., {et~al.} 2014, \apj, 797,
  122

\bibitem[{{Duchon}(1976)}]{Duchon1976}
{Duchon}, J. 1976, ESAIM: Mathematical Modelling and Numerical Analysis -
  Mod{\'e}lisation Math{\'e}matique et Analyse Num{\'e}rique, 10, 5

\bibitem[{{Fitzpatrick}(1999)}]{Fitzpatrick1999}
{Fitzpatrick}, E.~L. 1999, \pasp, 111, 63

\bibitem[{{Folatelli} {et~al.}(2010){Folatelli}, {Phillips}, {Burns},
  {Contreras}, {Hamuy}, {Freedman}, {Persson}, {Stritzinger}, {Suntzeff},
  {Krisciunas}, {Boldt}, {Gonz{\'a}lez}, {Krzeminski}, {Morrell}, {Roth},
  {Salgado}, {Madore}, {Murphy}, {Wyatt}, {Li}, {Filippenko}, \&
  {Miller}}]{Folatelli2010}
{Folatelli}, G., {Phillips}, M.~M., {Burns}, C.~R., {et~al.} 2010, \aj, 139,
  120

\bibitem[{{Garc{\'{\i}}a} {et~al.}(2014){Garc{\'{\i}}a}, {Ceillier},
  {Salabert}, {Mathur}, {van Saders}, {Pinsonneault}, {Ballot}, {Beck},
  {Bloemen}, {Campante}, {Davies}, {do Nascimento}, {Mathis}, {Metcalfe},
  {Nielsen}, {Su{\'a}rez}, {Chaplin}, {Jim{\'e}nez}, \& {Karoff}}]{Garcia2014}
{Garc{\'{\i}}a}, R.~A., {Ceillier}, T., {Salabert}, D., {et~al.} 2014, \aap,
  572, A34

\bibitem[{{Girardi} {et~al.}(2012){Girardi}, {Barbieri}, {Groenewegen},
  {Marigo}, {Bressan}, {Rocha-Pinto}, {Santiago}, {Camargo}, \& {da
  Costa}}]{Girardi2012}
{Girardi}, L., {Barbieri}, M., {Groenewegen}, M.~A.~T., {et~al.} 2012,
  {TRILEGAL, a TRIdimensional modeL of thE GALaxy: Status and Future}, ed.
  A.~{Miglio}, J.~{Montalb{\'a}n}, \& A.~{Noels}, 165

\bibitem[{{Hartman} {et~al.}(2008){Hartman}, {Gaudi}, {Holman}, {McLeod},
  {Stanek}, {Barranco}, {Pinsonneault}, \& {Kalirai}}]{Hartman2008}
{Hartman}, J.~D., {Gaudi}, B.~S., {Holman}, M.~J., {et~al.} 2008, \apj, 675,
  1254

\bibitem[{{Hartman} {et~al.}(2005){Hartman}, {Stanek}, {Gaudi}, {Holman}, \&
  {McLeod}}]{Hartman2005}
{Hartman}, J.~D., {Stanek}, K.~Z., {Gaudi}, B.~S., {Holman}, M.~J., \&
  {McLeod}, B.~A. 2005, \aj, 130, 2241

\bibitem[{{Hawley} {et~al.}(2014){Hawley}, {Davenport}, {Kowalski},
  {Wisniewski}, {Hebb}, {Deitrick}, \& {Hilton}}]{Hawley2014}
{Hawley}, S.~L., {Davenport}, J.~R.~A., {Kowalski}, A.~F., {et~al.} 2014, \apj,
  797, 121

\bibitem[{{Jofr{\'e}} \& {Weiss}(2011)}]{Jofre2011}
{Jofr{\'e}}, P., \& {Weiss}, A. 2011, \aap, 533, A59

\bibitem[{{Kaluzny} {et~al.}(1998){Kaluzny}, {Stanek}, {Krockenberger},
  {Sasselov}, {Tonry}, \& {Mateo}}]{Kaluzny1998}
{Kaluzny}, J., {Stanek}, K.~Z., {Krockenberger}, M., {et~al.} 1998, \aj, 115,
  1016

\bibitem[{{Kessler} {et~al.}(2015){Kessler}, {Marriner}, {Childress},
  {Covarrubias}, {D'Andrea}, {Finley}, {Fischer}, {Foley}, {Goldstein},
  {Gupta}, {Kuehn}, {Marcha}, {Nichol}, {Papadopoulos}, {Sako}, {Scolnic},
  {Smith}, {Sullivan}, {Wester}, {Yuan}, {Abbott}, {Abdalla}, {Allam},
  {Benoit-Levy}, {Bernstein}, {Bertin}, {Brooks}, {Carnero Rosell}, {Carrasco
  Kind}, {Castander}, {Crocce}, {da Costa}, {Desai}, {Diehl}, {Eifler}, {Fausti
  Neto}, {Flaugher}, {Frieman}, {Gruen}, {Gruendl}, {Honscheid}, {James},
  {Kuropatkin}, {Li}, {Maia}, {Marshall}, {Martini}, {Miller}, {Miquel},
  {Ogando}, {Plazas}, {Romer}, {Roodman}, {Sanchez}, {Sevilla-Noarbe}, {Smith},
  {Soares-Santos}, {Sobreira}, {Tarle}, {Thaler}, {Thomas}, {Tucker}, \&
  {Walker}}]{Kessler2015}
{Kessler}, R., {Marriner}, J., {Childress}, M., {et~al.} 2015, ArXiv e-prints,
  arXiv:1507.05137

\bibitem[{{Kov{\'a}cs} {et~al.}(2005){Kov{\'a}cs}, {Bakos}, \&
  {Noyes}}]{Kovacs2005}
{Kov{\'a}cs}, G., {Bakos}, G., \& {Noyes}, R.~W. 2005, \mnras, 356, 557

\bibitem[{{Kov{\'a}cs} {et~al.}(2002){Kov{\'a}cs}, {Zucker}, \&
  {Mazeh}}]{Kovacs2002}
{Kov{\'a}cs}, G., {Zucker}, S., \& {Mazeh}, T. 2002, \aap, 391, 369

\bibitem[{{Kowalski} {et~al.}(2009){Kowalski}, {Hawley}, {Hilton}, {Becker},
  {West}, {Bochanski}, \& {Sesar}}]{Kowalski2009}
{Kowalski}, A.~F., {Hawley}, S.~L., {Hilton}, E.~J., {et~al.} 2009, \aj, 138,
  633

\bibitem[{{Law} {et~al.}(2009){Law}, {Kulkarni}, {Dekany}, {Ofek}, {Quimby},
  {Nugent}, {Surace}, {Grillmair}, {Bloom}, {Kasliwal}, {Bildsten}, {Brown},
  {Cenko}, {Ciardi}, {Croner}, {Djorgovski}, {van Eyken}, {Filippenko}, {Fox},
  {Gal-Yam}, {Hale}, {Hamam}, {Helou}, {Henning}, {Howell}, {Jacobsen},
  {Laher}, {Mattingly}, {McKenna}, {Pickles}, {Poznanski}, {Rahmer}, {Rau},
  {Rosing}, {Shara}, {Smith}, {Starr}, {Sullivan}, {Velur}, {Walters}, \&
  {Zolkower}}]{Law2009}
{Law}, N.~M., {Kulkarni}, S.~R., {Dekany}, R.~G., {et~al.} 2009, \pasp, 121,
  1395

\bibitem[{{Law} {et~al.}(2013){Law}, {Carlberg}, {Salbi}, {Ngan}, {Ahmadi},
  {Steinbring}, {Murowinski}, {Sivanandam}, \& {Kerzendorf}}]{Law2013}
{Law}, N.~M., {Carlberg}, R., {Salbi}, P., {et~al.} 2013, \aj, 145, 58

\bibitem[{{Lawrence} {et~al.}(2009){Lawrence}, {Ashley}, {Hengst}, {Luong-van},
  {Storey}, {Yang}, {Zhou}, \& {Zhu}}]{Lawrence2009}
{Lawrence}, J.~S., {Ashley}, M.~C.~B., {Hengst}, S., {et~al.} 2009, Review of
  Scientific Instruments, 80, 064501

\bibitem[{{Lomb}(1976)}]{Lomb}
{Lomb}, N.~R. 1976, \apss, 39, 447

\bibitem[{{Lurie} {et~al.}(2015){Lurie}, {Davenport}, {Hawley}, {Wilkinson},
  {Wisniewski}, {Kowalski}, \& {Hebb}}]{Lurie2015}
{Lurie}, J.~C., {Davenport}, J.~R.~A., {Hawley}, S.~L., {et~al.} 2015, \apj,
  800, 95

\bibitem[{{Meng} {et~al.}(2013){Meng}, {Zhou}, {Zhang}, {Zhou}, {Wang}, {Ma},
  {Zhang}, {Fan}, \& {Zou}}]{Meng2013}
{Meng}, Z., {Zhou}, X., {Zhang}, H., {et~al.} 2013, \pasp, 125, 1015

\bibitem[{{Miller} {et~al.}(2008){Miller}, {Pennypacker}, \&
  {White}}]{Miller2008}
{Miller}, J.~P., {Pennypacker}, C.~R., \& {White}, G.~L. 2008, \pasp, 120, 449

\bibitem[{{O'Connell}(1951)}]{OConnell1951}
{O'Connell}, D.~J.~K. 1951, Publications of the Riverview College Observatory,
  2, 85

\bibitem[{{Oelkers} {et~al.}(2015){Oelkers}, {Macri}, {Wang}, {Ashley}, {Cui},
  {Feng}, {Gong}, {Lawrence}, {Qiang}, {Luong-Van}, {Pennypacker}, {Yang},
  {Yuan}, {York}, {Zhou}, \& {Zhu}}]{Oelkers2015}
{Oelkers}, R.~J., {Macri}, L.~M., {Wang}, L., {et~al.} 2015, \aj, 149, 50

\bibitem[{{Pecaut} \& {Mamajek}(2013)}]{Pecaut2013}
{Pecaut}, M.~J., \& {Mamajek}, E.~E. 2013, \apjs, 208, 9

\bibitem[{{Pepper} {et~al.}(2007){Pepper}, {Pogge}, {DePoy}, {Marshall},
  {Stanek}, {Stutz}, {Poindexter}, {Siverd}, {O'Brien}, {Trueblood}, \&
  {Trueblood}}]{Pepper2007}
{Pepper}, J., {Pogge}, R.~W., {DePoy}, D.~L., {et~al.} 2007, \pasp, 119, 923

\bibitem[{{Phillips}(1993)}]{Phillips1993}
{Phillips}, M.~M. 1993, \apjl, 413, L105

\bibitem[{{Pollacco} {et~al.}(2006){Pollacco}, {Skillen}, {Collier Cameron},
  {Christian}, {Hellier}, {Irwin}, {Lister}, {Street}, {West}, {Anderson},
  {Clarkson}, {Deeg}, {Enoch}, {Evans}, {Fitzsimmons}, {Haswell}, {Hodgkin},
  {Horne}, {Kane}, {Keenan}, {Maxted}, {Norton}, {Osborne}, {Parley}, {Ryans},
  {Smalley}, {Wheatley}, \& {Wilson}}]{Pollacco2006}
{Pollacco}, D.~L., {Skillen}, I., {Collier Cameron}, A., {et~al.} 2006, \pasp,
  118, 1407

\bibitem[{{Qian} {et~al.}(2014){Qian}, {Wang}, {Zhu}, {Snoonthornthum}, {Wang},
  {Zhao}, {Zhou}, {Liao}, \& {Liu}}]{Qian2014}
{Qian}, S.-B., {Wang}, J.-J., {Zhu}, L.-Y., {et~al.} 2014, \apjs, 212, 4

\bibitem[{{Scargle}(1982)}]{Scargle}
{Scargle}, J.~D. 1982, \apj, 263, 835

\bibitem[{{Schlafly} \& {Finkbeiner}(2011)}]{Schlafly2011}
{Schlafly}, E.~F., \& {Finkbeiner}, D.~P. 2011, \apj, 737, 103

\bibitem[{{Shappee} {et~al.}(2014){Shappee}, {Prieto}, {Grupe}, {Kochanek},
  {Stanek}, {De Rosa}, {Mathur}, {Zu}, {Peterson}, {Pogge}, {Komossa}, {Im},
  {Jencson}, {Holoien}, {Basu}, {Beacom}, {Szczygie{\l}}, {Brimacombe},
  {Adams}, {Campillay}, {Choi}, {Contreras}, {Dietrich}, {Dubberley},
  {Elphick}, {Foale}, {Giustini}, {Gonzalez}, {Hawkins}, {Howell}, {Hsiao},
  {Koss}, {Leighly}, {Morrell}, {Mudd}, {Mullins}, {Nugent}, {Parrent},
  {Phillips}, {Pojmanski}, {Rosing}, {Ross}, {Sand}, {Terndrup}, {Valenti},
  {Walker}, \& {Yoon}}]{Shappee2014}
{Shappee}, B.~J., {Prieto}, J.~L., {Grupe}, D., {et~al.} 2014, \apj, 788, 48

\bibitem[{{Skrutskie} {et~al.}(2006){Skrutskie}, {Cutri}, {Stiening},
  {Weinberg}, {Schneider}, {Carpenter}, {Beichman}, {Capps}, {Chester},
  {Elias}, {Huchra}, {Liebert}, {Lonsdale}, {Monet}, {Price}, {Seitzer},
  {Jarrett}, {Kirkpatrick}, {Gizis}, {Howard}, {Evans}, {Fowler}, {Fullmer},
  {Hurt}, {Light}, {Kopan}, {Marsh}, {McCallon}, {Tam}, {Van Dyk}, \&
  {Wheelock}}]{Skrutskie2006}
{Skrutskie}, M.~F., {Cutri}, R.~M., {Stiening}, R., {et~al.} 2006, \aj, 131,
  1163

\bibitem[{{Stetson}(1987)}]{Stetson1987}
{Stetson}, P.~B. 1987, \pasp, 99, 191

\bibitem[{{Stetson}(1996)}]{Stetson1996}
---. 1996, \pasp, 108, 851

\bibitem[{{Udalski} {et~al.}(1994){Udalski}, {Szymanski}, {Stanek}, {Kaluzny},
  {Kubiak}, {Mateo}, {Krzeminski}, {Paczynski}, \& {Venkat}}]{Udalski1994}
{Udalski}, A., {Szymanski}, M., {Stanek}, K.~Z., {et~al.} 1994, \actaa, 44, 165

\bibitem[{{Wang, L.} {et~al.}(2011){Wang, L.}, {Macri}, {Krisciunas}, {Wang},
  {Ashley}, {Cui}, {Feng}, {Gong}, {Lawrence}, {Liu}, {Luong-Van},
  {Pennypacker}, {Shang}, {Storey}, {Yang}, {Yang}, {Yuan}, {York}, {Zhou},
  {Zhu}, \& {Zhu}}]{Wang2011}
{Wang, L.}, {Macri}, L.~M., {Krisciunas}, K., {et~al.} 2011, \aj, 142, 155

\bibitem[{{Wang, L.} {et~al.}(2013){Wang, L.}, {Macri}, {Wang}, {Ashley},
  {Cui}, {Feng}, {Gong}, {Lawrence}, {Liu}, {Luong-Van}, {Pennypacker},
  {Shang}, {Storey}, {Yang}, {Yang}, {Yuan}, {York}, {Zhou}, {Zhu}, \&
  {Zhu}}]{Wang2013}
{Wang, L.}, {Macri}, L.~M., {Wang}, L., {et~al.} 2013, \aj, 146, 139

\bibitem[{{Wang, S.} {et~al.}(2012){Wang, S.}, {Zhou}, {Zhang}, {Zhou}, {Ma},
  {Fan}, {Zou}, {Meng}, \& {Yang}}]{WangS2012}
{Wang, S.}, {Zhou}, X., {Zhang}, H., {et~al.} 2012, \pasp, 124, 1167

\bibitem[{{Wang, S.} {et~al.}(2014){Wang, S.}, {Zhang}, {Zhou}, {Zhou}, {Yang},
  {Wang}, {Bayliss}, {Zhou}, {Ashley}, {Fan}, {Feng}, {Gong}, {Lawrence},
  {Liu}, {Liu}, {Luong-Van}, {Ma}, {Meng}, {Storey}, {Wittenmyer}, {Wu}, {Yan},
  {Yang}, {Yang}, {Yang}, {Yuan}, {Zhang}, {Zhu}, \& {Zou}}]{WangS2014}
{Wang, S.}, {Zhang}, H., {Zhou}, J.-L., {et~al.} 2014, \apjs, 211, 26

\bibitem[{{Wang, S.} {et~al.}(2015){Wang, S.}, {Zhang}, {Zhou}, {Zhou}, {Fu},
  {Yang}, {Liu}, {Xie}, {Wang}, {Wang}, {Wittenmyer}, {Ashley}, {Feng}, {Gong},
  {Lawrence}, {Liu}, {Luong-Van}, {Ma}, {Peng}, {Storey}, {Wu}, {Yan}, {Yang},
  {Yang}, {Yuan}, {Zhang}, {Zhang}, {Zhu}, \& {Zou}}]{WangS2015}
{Wang, S.}, {Zhang}, H., {Zhou}, X., {et~al.} 2015, \apjs, 218, 20

\bibitem[{{Wang, S.-H.} {et~al.}(2014){Wang, S.-H.}, {Zhou}, {Zhang}, {Zhou},
  {Liu}, {Meng}, {Ma}, {Zhang}, {Fan}, \& {Zou}}]{WangS2014b}
{Wang, S.-H.}, {Zhou}, X., {Zhang}, H., {et~al.} 2014, Research in Astronomy
  and Astrophysics, 14, 345

\bibitem[{{Yang} {et~al.}(2015){Yang}, {Zhang}, {Wang}, {Zhou}, {Zhou}, {Wang},
  {Wang}, {Wittenmyer}, {Liu}, {Meng}, {Ashley}, {Storey}, {Bayliss}, {Tinney},
  {Wang}, {Wu}, {Liang}, {Yu}, {Fan}, {Feng}, {Gong}, {Lawrence}, {Liu},
  {Luong-Van}, {Ma}, {Wu}, {Yan}, {Yang}, {Yang}, {Yuan}, {Zhang}, {Zhu}, \&
  {Zou}}]{Yang2015}
{Yang}, M., {Zhang}, H., {Wang}, S., {et~al.} 2015, \apjs, 217, 28

\bibitem[{{York} {et~al.}(2000){York}, {Adelman}, {Anderson}, {Anderson},
  {Annis}, {Bahcall}, {Bakken}, {Barkhouser}, {Bastian}, {Berman}, {Boroski},
  {Bracker}, {Briegel}, {Briggs}, {Brinkmann}, {Brunner}, {Burles}, {Carey},
  {Carr}, {Castander}, {Chen}, {Colestock}, {Connolly}, {Crocker}, {Csabai},
  {Czarapata}, {Davis}, {Doi}, {Dombeck}, {Eisenstein}, {Ellman}, {Elms},
  {Evans}, {Fan}, {Federwitz}, {Fiscelli}, {Friedman}, {Frieman}, {Fukugita},
  {Gillespie}, {Gunn}, {Gurbani}, {de Haas}, {Haldeman}, {Harris}, {Hayes},
  {Heckman}, {Hennessy}, {Hindsley}, {Holm}, {Holmgren}, {Huang}, {Hull},
  {Husby}, {Ichikawa}, {Ichikawa}, {Ivezi{\'c}}, {Kent}, {Kim}, {Kinney},
  {Klaene}, {Kleinman}, {Kleinman}, {Knapp}, {Korienek}, {Kron}, {Kunszt},
  {Lamb}, {Lee}, {Leger}, {Limmongkol}, {Lindenmeyer}, {Long}, {Loomis},
  {Loveday}, {Lucinio}, {Lupton}, {MacKinnon}, {Mannery}, {Mantsch}, {Margon},
  {McGehee}, {McKay}, {Meiksin}, {Merelli}, {Monet}, {Munn}, {Narayanan},
  {Nash}, {Neilsen}, {Neswold}, {Newberg}, {Nichol}, {Nicinski}, {Nonino},
  {Okada}, {Okamura}, {Ostriker}, {Owen}, {Pauls}, {Peoples}, {Peterson},
  {Petravick}, {Pier}, {Pope}, {Pordes}, {Prosapio}, {Rechenmacher}, {Quinn},
  {Richards}, {Richmond}, {Rivetta}, {Rockosi}, {Ruthmansdorfer}, {Sandford},
  {Schlegel}, {Schneider}, {Sekiguchi}, {Sergey}, {Shimasaku}, {Siegmund},
  {Smee}, {Smith}, {Snedden}, {Stone}, {Stoughton}, {Strauss}, {Stubbs},
  {SubbaRao}, {Szalay}, {Szapudi}, {Szokoly}, {Thakar}, {Tremonti}, {Tucker},
  {Uomoto}, {Vanden Berk}, {Vogeley}, {Waddell}, {Wang}, {Watanabe},
  {Weinberg}, {Yanny}, {Yasuda}, \& {SDSS Collaboration}}]{York2000}
{York}, D.~G., {Adelman}, J., {Anderson}, Jr., J.~E., {et~al.} 2000, \aj, 120,
  1579

\bibitem[{{Young}(1967)}]{Young1967}
{Young}, A.~T. 1967, \aj, 72, 747

\bibitem[{{Yuan} {et~al.}(2008){Yuan}, {Cui}, {Liu}, {Zhai}, {Gong}, {Zhang},
  {Xia}, {Hu}, {Lawrence}, {Yan}, {Storey}, {Wang}, {Feng}, {Ashley}, {Zhou},
  {Jiang}, \& {Zhu}}]{Yuan2008}
{Yuan}, X., {Cui}, X., {Liu}, G., {et~al.} 2008, in Society of Photo-Optical
  Instrumentation Engineers (SPIE) Conference Series, Vol. 7012, Society of
  Photo-Optical Instrumentation Engineers (SPIE) Conference Series

\bibitem[{{Zhou} {et~al.}(2010{\natexlab{a}}){Zhou}, {Wu}, {Jiang}, {Cui},
  {Feng}, {Gong}, {Hu}, {Li}, {Liu}, {Ma}, {Wang}, {Wang}, {Wu}, {Xia}, {Yan},
  {Yuan}, {Zhai}, {Zhang}, \& {Zhu}}]{Zhou2010a}
{Zhou}, X., {Wu}, Z.-Y., {Jiang}, Z.-J., {et~al.} 2010{\natexlab{a}}, Research
  in Astronomy and Astrophysics, 10, 279

\bibitem[{{Zhou} {et~al.}(2010{\natexlab{b}}){Zhou}, {Fan}, {Jiang}, {Ashley},
  {Cui}, {Feng}, {Gong}, {Hu}, {Kulesa}, {Lawrence}, {Liu}, {Luong-Van}, {Ma},
  {Moore}, {Qin}, {Shang}, {Storey}, {Sun}, {Travouillon}, {Walker}, {Wang},
  {Wang}, {Wu}, {Wu}, {Xia}, {Yan}, {Yang}, {Yang}, {Yuan}, {York}, {Zhang}, \&
  {Zhu}}]{Zhou2010}
{Zhou}, X., {Fan}, Z., {Jiang}, Z., {et~al.} 2010{\natexlab{b}}, \pasp, 122,
  347

\bibitem[{{Zhou} {et~al.}(2013){Zhou}, {Ashley}, {Cui}, {Feng}, {Gong}, {Hu},
  {Jiang}, {Kulesa}, {Lawrence}, {Liu}, {Luong-Van}, {Ma}, {Macri}, {Meng},
  {Moore}, {Qin}, {Shang}, {Storey}, {Sun}, {Travouillon}, {Walker}, {Wang},
  {Wang}, {Wang}, {Wang}, {Wu}, {Wu}, {Xia}, {Yan}, {Yang}, {Yang}, {Yao},
  {Yuan}, {York}, {Zhang}, {Zhang}, {Zhou}, {Zhu}, \& {Zou}}]{Zhou2013}
{Zhou}, X., {Ashley}, M.~C.~B., {Cui}, X., {et~al.} 2013, in IAU Symposium,
  Vol. 288, IAU Symposium, ed. M.~G. {Burton}, X.~{Cui}, \& N.~F.~H. {Tothill},
  231--238

\bibitem[{{Zou} {et~al.}(2010){Zou}, {Zhou}, {Jiang}, {Ashley}, {Cui}, {Feng},
  {Gong}, {Hu}, {Kulesa}, {Lawrence}, {Liu}, {Luong-Van}, {Ma}, {Moore},
  {Pennypacker}, {Qin}, {Shang}, {Storey}, {Sun}, {Travouillon}, {Walker},
  {Wang}, {Wang}, {Wu}, {Wu}, {Xia}, {Yan}, {Yang}, {Yang}, {Yao}, {Yuan},
  {York}, {Zhang}, \& {Zhu}}]{Zou2010}
{Zou}, H., {Zhou}, X., {Jiang}, Z., {et~al.} 2010, \aj, 140, 602

\end{thebibliography}

\begin{deluxetable*}{cccccccccc}
\tabletypesize{\tiny}
\tablewidth{0pt}
\tablecaption{Identified Candidate Stellar Flares \label{tb:tran}}
\tablehead{\colhead{ID} &  \colhead{R.A.} & \colhead{Dec.} & \colhead{Filter} & \colhead{K/M dwarf}& \colhead{MJD-2454500} & \colhead{Length [d]} & \colhead{Amplitude [mag]}& \multicolumn{2}{c}{Comment*}}
\startdata
CSTARJ111143.79-875135 & 11:11:43.79 & -87:51:35 & i & K5V & 887.181580 & 0.350 & 0.027 &            0 & \\
CSTARJ100426.56-883937 & 10:04:26.56 & -88:39:37 & i & \nd & 881.362122 & 0.065 & 0.016 &            0 &  \\
CSTARJ104851.25-882931 & 10:48:51.25 & -88:29:31 & i & \nd & 859.818665 & 0.102 & 0.096 &           0 &  \\
CSTARJ112329.83-891523 & 11:23:29.83 & -89:15:23 & i & \nd & 858.012878 & 0.022 & 0.336 &            0 & \\
CSTARJ150830.36-885721 & 15:08:30.36 & -88:57:21 & i & \nd & 832.369446 & 0.324 & 0.015 &            0 & \\
CSTARJ115040.50-892747 & 11:50:40.50 & -89:27:47 & i & K6V & 827.375916 & 0.777 & 0.014 &            0 &  \\
CSTARJ064616.70-874825 & 06:46:16.70 & -87:48:25 & i & \nd & 847.265564 & 0.023 & 0.298 &            0 & \\
CSTARJ062327.70-875637 & 06:23:27.70 & -87:56:37 & i & \nd & 886.023499 & 0.310 & 0.057 &           0 &  \\
CSTARJ210848.50-892830 & 21:08:48.50 & -89:28:30 & i & \nd & 865.311279 & 0.022 & 0.574 &            0  & \\
CSTARJ034834.23-882808 & 03:48:34.23 & -88:28:08 & i & \nd & 861.093262 & 0.490 & 0.045 &            0  & \\
CSTARJ021530.10-873840 & 02:15:30.10 & -87:38:40 & g & \nd & 468.947968 & 0.028 & 0.081 &  4 & \\
CSTARJ054239.09-872933 & 05:42:39.09 & -87:29:33 & g & \nd & 476.092010 & 0.270 & 0.033 &  4 & \\
CSTARJ100426.56-883937 & 10:04:26.56 & -88:39:37 & g & \nd & 478.015503 & 0.033 & 0.032 &  4 & \\
CSTARJ153412.75-881014 & 15:34:12.75 & -88:10:14 & g & \nd & 472.540314 & 0.260 & 0.260 &      1 & \\
CSTARJ060139.40-880138 & 06:01:39.40 & -88:01:38 & g & \nd & 481.687744 & 0.410 & 0.049 &   2 & \\
CSTARJ060745.20-882219 & 06:07:45.20 & -88:22:19 & g & M2V & 479.106659 & 0.032 & 0.031 &   4 & \\
CSTARJ203720.10-880633 & 20:37:20.10 & -88:06:33 & g & \nd & 471.879669 & 0.120 & 0.034 &   4 & \\
CSTARJ172731.23-885235 & 17:27:31.23 & -88:52:35 & g &\nd & 477.950073 & 0.140 & 0.018 &   4 & \\
CSTARJ122630.23-882031 & 12:26:30.23 & -88:20:31 & g & \nd & 449.685028 & 0.248 & 0.665 &   3 & \\
CSTARJ100731.56-884333 & 10:07:31.56 & -88:43:33 & g & \nd & 471.660156 & 0.230 & 0.320 &   4 & \\
CSTARJ100737.56-880919 & 10:07:37.56 & -88:09:19 & r & \nd & 506.169220 & 0.120 & 0.045 & 3 & \\
CSTARJ090621.10-882040 & 09:06:21.10 & -88:20:40 & r & M0V & 490.181580 & 0.230 & 0.219 & 3 & \\
CSTARJ042931.75-893647 & 04:29:31.75 & -89:36:47 & r & \nd & 481.773285 & 0.151 & 0.779 &   2 & \\
CSTARJ010728.36-885510 & 01:07:28.36 & -88:55:10 & r & \nd & 534.605530 & 0.215 & 0.151 & 3 & \\
CSTARJ022220.06-875619 & 02:22:20.06 & -87:56:19 & r & \nd & 485.373688 & 0.290 & 0.283 & 3 & \\
CSTARJ053807.80-875538 & 05:38:07.80 & -87:55:38 & r & \nd & 506.498016 & 0.370 & 0.389 & 2 & 3 \\
CSTARJ090738.25-884334 & 09:07:38.25 & -88:43:34 & r & \nd & 527.660156 & 0.076 & 0.564 & 2 & 3 \\
CSTARJ210122.50-874600 & 21:01:22.50 & -87:46:00 & r & \nd & 531.661865 & 0.320 & 0.162 & 2 & 3 \\
CSTARJ141243.70-883232 & 14:12:43.70 & -88:32:32 & r & M7V & 506.367676 & 0.399 & 0.154 & 3 & \\
\enddata
\tablecomments{*: The comments have flags from 0-4 as follows: Only the \textit{i} band was observed in 2010 and would not have any counterpart bands (0); the candidate flare star had no matched star in the other band (2); the candidate flare was observed outside of simultaneous observations, MJD 54955-985 (3); the candidate flare was observed during a data drop out in the counterpart band (4).}
\end{deluxetable*}

\end{document}